\documentclass[11pt]{article}
\usepackage{amsmath}
\usepackage{amssymb}
\usepackage{graphicx}
\usepackage{hyperref} 
\usepackage{cite}
\usepackage{bigstrut}
\usepackage{gensymb}
\usepackage{booktabs}
\usepackage{multirow}

\makeatletter

\@addtoreset{equation}{section}

\allowdisplaybreaks[2]

\newcommand\Refr[1]     {Ref.\,\cite{#1}}

\newcommand\refr[1]     {ref.\,\cite{#1}}
\newcommand\refrs[1]    {refs.\,\cite{#1}}

\newcommand\Eqn[1]     {Eq.\,(\ref{#1})}

\newcommand\eqn[1]     {eq.\,(\ref{#1})}
\newcommand\eqns[2]    {eqs.\,(\ref{#1}) and~(\ref{#2})}
\newcommand\eqnss[2]   {eqs.\,(\ref{#1})--(\ref{#2})}

\newcommand\Fig[1]     {Figure~{\ref{#1}}}
\newcommand\fig[1]     {figure~{\ref{#1}}}

\newcommand\sect[1]    {section~{\ref{#1}}}

\newcommand\appx[1]     {appendix~\ref{#1}}

\newcommand\tab[1]     {table~\ref{#1}}

\def\beq{\begin{equation}}
\def\eeq{\end{equation}}
\def\bsp#1\esp{\begin{split}#1\end{split}}
\def\bal#1\eal{\begin{align}#1\end{align}}
\newcommand\nt         {\notag}

\newcommand\tot		  {\rm{t}}
\newcommand\tsigT      {\sigma_{\tot}}
\newcommand\tsig[2]    {\sigma^{\rm{#1}}_{#2}}
\newcommand\dsig[2]    {\rd\sigma^{{\rm #1}}_{#2}}

\newcommand{\rd}       {{\rm{d}}}

\newcommand\as  	       {\ensuremath{\alpha_{\rm{S}}}}
\newcommand\Oa[1]      {\ensuremath{\mathcal O(\as^{#1})}}

\newcommand{\CF}       {C_{\rm{F}}}
\newcommand{\CA}       {C_{\rm{A}}}
\newcommand{\TR}       {T_{\rm{R}}}
\newcommand{\Nc}       {N_{\rm{c}}}
\newcommand{\Nf}       {\ensuremath{n_{\rm{f}}}}

\newcommand{\xiRsq}      {\frac{\mu^2}{Q^2}}
\newcommand{\colorfulNNLO}{{CoLoRFulNNLO}}
\newcommand\MSbar	  {\ensuremath{\overline{\rm{MS}}}}

\begin{document}


\begin{titlepage}
\renewcommand{\thefootnote}{\arabic{footnote}}
\begin{flushright}
arXiv:1708.xxxxx
\end{flushright}
\par \vspace{10mm}

\begin{center}
{\Large \bf Energy-energy correlation in electron-positron }
\\[0.25cm]
{\Large \bf annihilation at NNLL+NNLO accuracy}
\end{center}
\par \vspace{2mm}
\begin{center}
{\bf Zolt\'an Tulip\'ant},
{\bf Adam Kardos} and 
{\bf G\'abor Somogyi}\\
\vspace{5mm}

MTA-DE Particle Physics Research Group, University of Debrecen, 4010 Debrecen,
\\ PO Box 105, Hungary

\vspace{5mm}

\end{center}

\par \vspace{2mm}
\begin{center} {\large \bf Abstract} \end{center}
\begin{quote}
\pretolerance 10000

We present the computation of energy-energy correlation in $e^+e^-$ collisions 
in the back-to-back region at next-to-next-to-leading logarithmic accuracy 
matched with the next-to-next-to-leading order perturbative prediction. We study 
the effect of the fixed higher-order corrections in a comparison of our results to 
LEP and SLC data. The next-to-next-to-leading order correction has a sizable impact 
on the extracted value of $\as(M_Z$), hence its inclusion is mandatory for a precise 
measurement of the strong coupling using energy-energy correlation.
\end{quote}

\vspace*{\fill}
\begin{flushleft}
August 2017

\end{flushleft}
\end{titlepage}

\setcounter{footnote}{0}
\renewcommand{\thefootnote}{\arabic{footnote}}


\section{Introduction}
\label{sec:intro}

Precision measurements of event shape distributions in $e^+e^-$ annihilation 
have provided detailed experimental tests of QCD and remain one of the most 
precise tools used for extracting the strong coupling $\as$ from data. Quantities 
related to three-jet events are particularly well suited for this task, since the 
deviation from simple two-jet configurations is directly proportional to $\as$. 
Furthermore, since the strong interactions occur only in the final state, 
non-perturbative QCD corrections are restricted to hadronization and power 
corrections. These may either be extracted from data by comparison to predictions 
by Monte Carlo simulations or computed using analytic models. Hence the precision 
of the theoretical computation is limited mainly by the accuracy of the perturbative 
expansion in $\as$.

In this regard, the state of the art currently includes exact fixed-order 
next-to-next-to-leading order (NNLO) corrections for the six standard three-jet 
event shapes of thrust, heavy jet mass, total and wide jet broadening, $C$-parameter 
and the two-to-three jet transition variable $y_{23}$ \cite{GehrmannDeRidder:2007hr,
Weinzierl:2009ms,DelDuca:2016ily} as well as jet cone energy fraction 
\cite{DelDuca:2016ily}, oblateness and energy-energy correlation \cite{DelDuca:2016csb}.

However, fixed-order predictions have a limited kinematical range of 
applicability. For example when the two-jet limit is approached multiple emissions of 
soft and collinear gluons give rise to large logarithmic corrections that invalidate 
the use of fixed-order perturbation theory. In order to obtain a description appropriate 
to this limit, the logarithms must be resummed to all orders. For three-jet event shapes 
such logarithmically enhanced terms can be resummed at next-to-next-to-leading logarithmic 
(NNLL) accuracy \cite{deFlorian:2004mp,Becher:2008cf,Chien:2010kc,Monni:2011gb,Alioli:2012fc,
Becher:2012qc,Banfi:2014sua} and even at next-to-next-to-next-to-leading logarithmic (N${}^3$LL) 
accuracy for thrust \cite{Abbate:2010xh} and the $C$-parameter \cite{Hoang:2014wka}. 
A prediction incorporating the complete perturbative knowledge about the observable can 
be derived by matching the fixed-order and resummed calculations. 

For the standard event shapes of thrust, heavy jet mass, total and wide jet broadening, 
$C$-parameter and $y_{23}$, NNLO predictions matched to NLL resummation were presented
in \refr{Gehrmann:2008kh}. Predictions at NNLO matched to N${}^3$LL resummation are also 
known for thrust \cite{Becher:2008cf,Abbate:2010xh} and the $C$-parameter \cite{Hoang:2014wka}. 

In this paper we consider the energy-energy correlation (EEC) in $e^+e^-$ annihilation 
and present NNLO predictions matched to NNLL resummation for the back-to-back region.
EEC was the first event shape for which a complete NNLL resummation was performed 
\cite{deFlorian:2004mp} while the fixed-order NNLO corrections to this observable were 
computed recently in \refr{DelDuca:2016csb}. We also investigate the numerical impact of our 
results and perform a comparison of the most accurate theoretical prediction with precise 
OPAL \cite{Acton:1993zh} and SLD \cite{Abe:1994mf} data.

The paper is organized as follows. In \sect{sec:theory} we review the ingredients 
of our calculation, i.e., the fixed-order result as well as the resummation formalism. 
The matching of the NNLO predictions to the NNLL resummation is not entirely 
straightforward and we devote \sect{sec:matching} to a careful discussion of our 
procedure. We compare our results to LEP and SLC measurements in \sect{sec:pheno} and 
in particular perform a fit to OPAL and SLD data. Finally, in \sect{sec:concl} we draw 
our conclusions.


\section{Fixed-order and resummed predictions}
\label{sec:theory}



EEC is the normalized energy-weighted cross section defined in terms of the angle between 
two particles $i$ and $j$ in an event \cite{Basham:1978bw}:
\beq
\frac{1}{\tsigT} \frac{\rd \Sigma(\chi)}{\rd \cos \chi} \equiv
	\frac{1}{\tsigT} \int \sum_{i,j} \frac{E_i E_j}{Q^2}
	\dsig{}{e^+e^- \to\, i j + X} \delta(\cos\chi + \cos\theta_{ij})\,,
\label{eq:dEEC-def}
\eeq
where $E_i$ and $E_j$ are particle energies, $Q$ is the center-of-mass energy, 
$\theta_{ij} = \pi - \chi$ is the angle between the two particles and $\tsigT$ 
is the total hadronic cross section. Notice that the back-to-back region, $\theta_{ij}\to \pi$ corresponds to $\chi \to 0$, while the normalization ensures that the integral of 
the EEC distribution from $\chi = 0\degree$ to $\chi = 180\degree$ is unity.

The fixed-order prediction for EEC has been known in QCD perturbation theory up to 
NLO accuracy for some time \cite{Richards:1983sr,Ellis:1983fg,Richards:1982te,Chao:1982wb,
Schneider:1983iu,Ali:1982ub,Kunszt:1989km,Falck:1988gb,Glover:1994vz,Clay:1995sd,Kramer:1996qr} 
and has been computed at NNLO accuracy recently in \refr{DelDuca:2016csb} 
using the \colorfulNNLO{} method \cite{Somogyi:2006da,Somogyi:2006db,DelDuca:2016ily}. 
At the renormalization scale $\mu$\footnote{We use the \MSbar{} renormalization scheme 
throughout the paper with the number of light quark flavors set to $\Nf=5$. Furthermore, we 
use the two-loop running of $\as$ for all predictions that incorporate a fixed-order NLO result, 
while predictions involving a fixed-order NNLO result are obtained using three-loop running.} 
the result can be written as:
\beq
\bigg[\frac{1}{\tsigT} \frac{\rd \Sigma(\chi,\mu)}{\rd \cos \chi}\bigg]_{\mathrm{f.o.}} = 
	\frac{\as(\mu)}{2\pi} \frac{\rd \bar{A}(\chi,\mu)}{\rd \cos \chi } 
	+
	\left(\frac{\as(\mu)}{2\pi}\right)^2 \frac{\rd \bar{B}(\chi,\mu) }{\rd \cos \chi }
	+
	\left(\frac{\as(\mu)}{2\pi}\right)^3 \frac{\rd \bar{C}(\chi,\mu)}{\rd \cos \chi }
	+
	\Oa{4}\,,
\label{eq:dEEC-fo}
\eeq
where $\bar{A}$, $\bar{B}$ and $\bar{C}$ are the perturbative coefficients at LO, NLO 
and NNLO, normalized to the total hadronic cross section. In practice, our numerical 
program computes this distribution normalized to $\tsig{}{0}$, the LO cross section for 
$e^+e^- \to \mbox{hadrons}$ and at the fixed scale of $\mu = Q$,
\beq
\bigg[\frac{1}{\tsig{}{0}} \frac{\rd \Sigma(\chi,Q)}{\rd \cos \chi}\bigg]_{\mathrm{f.o.}} = 
	\frac{\as(Q)}{2\pi} \frac{\rd A(\chi)}{\rd \cos \chi } 
	+
	\left(\frac{\as(Q)}{2\pi}\right)^2 \frac{\rd B(\chi) }{\rd \cos \chi }
	+
	\left(\frac{\as(Q)}{2\pi}\right)^3 \frac{\rd C(\chi)}{\rd \cos \chi }
	+
	\Oa{4}\,.
\label{eq:dEEC-fo-sigma0}
\eeq
At the default renormalization scale, the distribution normalized to the total hadronic 
cross section can be obtained from the expansion in \eqn{eq:dEEC-fo-sigma0} by multiplying 
with
\beq
\frac{\tsig{}{0}}{\tsigT} = 
	1 - \frac{\as(Q)}{2\pi} A_{\tot} + \left(\frac{\as(Q)}{2\pi}\right)^2 
	\left(A_{\tot}^2 - B_{\tot}\right) + \Oa{3}\,,
\eeq
where
\beq
A_{\tot} = \frac{3}{2}\CF 
\quad\mbox{and}\quad
B_{\tot} = \CF \left[\left(\frac{123}{8} - 11 \zeta_3\right) \CA
	- \frac{3}{8} \CF - \left(\frac{11}{2} - 4 \zeta_3\right) \Nf \TR \right]\,,
\eeq
with the color factors
\beq
\CA = 2\Nc\TR\,,
\qquad
\CF = \frac{\Nc^2-1}{\Nc}\TR
\qquad\mbox{and}\qquad 
\TR = \frac{1}{2}\,.
\eeq
Scale dependence can be restored using the renormalization group equation,
\bal
\mu^2\frac{\rd}{\rd \mu^2} \frac{\as(\mu)}{4\pi} &= 
	- \beta_0\left(\frac{\as(\mu)}{4\pi}\right)^2
	- \beta_1\left(\frac{\as(\mu)}{4\pi}\right)^3
	- \beta_2\left(\frac{\as(\mu)}{4\pi}\right)^4
	+ \Oa{5}\,,
\label{eq:RGE}\\
\beta_0 &= \frac{11\CA}{3} - \frac{4\Nf \TR}{3}\,,
\nt\\
\beta_1 &= \frac{34}{3} \CA^2 - \frac{20}{3} \CA \TR \Nf - 4 \CF \TR \Nf\,,
\nt\\
\beta_2 &= \frac{2857}{54} \CA^3
	- \left(\frac{1415}{27} \CA^2 + \frac{205}{9} \CA \CF - 2 \CF^2\right) \TR \Nf
	+ \left(\frac{158}{27} \CA + \frac{44}{9} \CF\right) \TR^2 \Nf^2
\nt\,,
\eal
and one finds
\beq
\bsp
\bar{A}(\chi,\mu) &= A(\chi)\,,
\\
\bar{B}(\chi,\mu) &= B(\chi) + \left(\frac{1}{2}\beta_0 \ln\xiRsq 
	- A_{\tot}\right) A(\chi)\,,
\\
\bar{C}(\chi,\mu) &= C(\chi) + \left(\beta_0 \ln\xiRsq - A_{\tot}\right) B(\chi)
\\ &
	+ \left(\frac{1}{4} \beta_1 \ln\xiRsq 
	+ \frac{1}{4}\beta_0^2 \ln^2\xiRsq
	-  A_{\tot} \beta_0 \ln\xiRsq 
	+ A_{\tot}^2 - B_{\tot} \right) A(\chi)\,.
\esp
\eeq
Finally, using three-loop running the scale dependence of the strong coupling is given 
by 
\beq
\as(\mu) = \as^{(1)}(\mu) + \as^{(2)}(\mu) + \as^{(3)}(\mu)\,,
\eeq
where $\as^{(1)}$, $\as^{(2)}$ and $\as^{(3)}$ represent the one-, two- and three-loop 
contributions:
\bal
\as^{(1)}(\mu) &= 
	\frac{\as(Q)}{1 + \as(Q) \frac{\beta_0}{4\pi} \ln\xiRsq}\,,
\\	
\as^{(2)}(\mu) &= 
	- (\as^{(1)}(\mu))^2 \frac{\beta_1}{4\pi\beta_0} \ln K(\mu)\,,
\nt\\
\as^{(3)}(\mu) &=
	(\as^{(1)}(\mu))^3 \left[\frac{\beta_1^2}{(4\pi)^2\beta_0^2}\ln K(\mu) 
		(\ln K(\mu) - 1)
		- \left(\frac{\beta_1^2}{(4\pi)^2\beta_0^2} - \frac{\beta_2}{(4\pi)^2\beta_0}\right)
		(1 - K(\mu))\right]\,,
\nt\eal
with
\beq
K(\mu) = \frac{\as(Q)}{\as^{(1)}(\mu)}\,.
\eeq

\begin{figure}[t]
\centering
\includegraphics[width=0.6\linewidth]{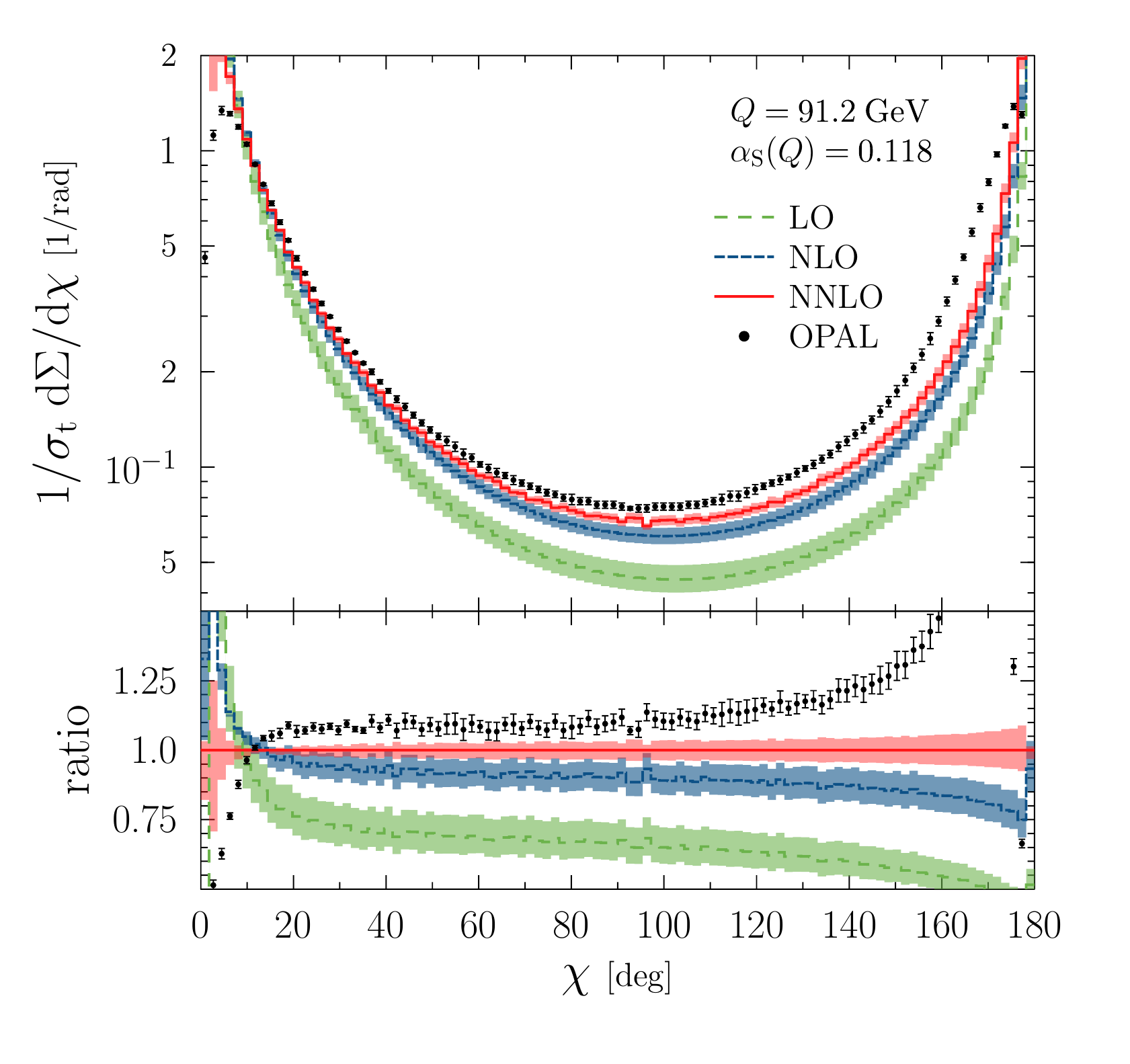}
\caption{Fixed-order predictions for EEC at LO, NLO and NNLO accuracy. The bottom 
panel shows the ratio of the data and the perturbative predictions at each order to 
the NNLO result. The bands represent the effects of varying the renormalization scale 
by a factor of two around the default scale of $\mu = Q$. OPAL data is also shown. }
\label{fig:dEEC-fo}
\end{figure}
The physical predictions for EEC up to NNLO accuracy are presented in \fig{fig:dEEC-fo} 
where the data measured by the OPAL collaboration is also shown. The bands represent 
the effect of varying the renormalization scale by a factor of two around its central 
value of $\mu = Q$ in both directions. Including the higher order corrections reduces 
the discrepancy between the predictions and data, although sizable differences remain. 
However, examining the region of intermediate $\chi$ ($\chi \gtrsim 30\degree$) i.e., 
the region of validity of the fixed-order expansion, we observe that the LO scale variation 
band does not overlap with the NLO one, while the overlap of the NLO band with the NNLO 
one is marginal up to around $\chi \sim 60\degree$, beyond which the two bands no longer 
touch. This behavior indicates that up to NLO the customary prescription for scale variation 
is not a reliable estimate of the size of the higher order corrections and casts some doubts 
also on the reliability of the NNLO band to estimate the perturbative uncertainty of the 
calculation. This phenomenon, however, is not unique to EEC and in fact very similar comments 
apply also to other three-jet event shapes in $e^+e^-$ annihilation 
\cite{GehrmannDeRidder:2007hr,Weinzierl:2009ms,DelDuca:2016ily}. Nevertheless, one could 
argue that a more realistic estimate of the perturbative uncertainty could be obtained by 
considering a wider range for scale variation, see \Refr{Bagnaschi:2014wea} for a careful 
discussion. This observation could explain, at least partially, the difference between the 
NNLO predictions and experimental data.

The fixed-order predictions clearly diverge for both small and large values 
of $\chi$. As discussed above, this is the result of large logarithmic contributions of 
infrared origin. Concentrating on the back-to-back region ($\theta_{ij}\to \pi$, i.e., 
$\chi \to 0$), these contributions take the form $\as^n \ln^{2n-1} y$, where
\beq
y = \sin^2 \frac{\chi}{2}\,.
\label{eq:y-def}
\eeq
As $y$ decreases the logarithms become large and invalidate the use of the fixed-order 
perturbative expansion.
In order to obtain a description of EEC in the small angle limit, these logarithmic 
contributions must be resummed to all orders. The appropriate resummation formalism has been 
developed in \refrs{Collins:1981uk,Collins:1981va,Collins:1985xx,Kodaira:1981nh,Kodaira:1982az}
and the coefficients which control this resummation are known completely at NNLL accuracy 
\cite{deFlorian:2004mp}\footnote{Note that the NNLL $A^{(3)}$ coefficient in 
\refr{deFlorian:2004mp} is incomplete. The full coefficient has been derived in 
\refr{Becher:2010tm}. (See also \refr{Banfi:2014sua}.)}. 
At a center-of-mass energy of $Q$ and renormalization scale $\mu$ the resummed 
prediction can be written as
\beq
\bigg[\frac{1}{\tsigT}\frac{\rd \Sigma(\chi,\mu)}{\rd \cos \chi}\bigg]_{\mathrm{res.}} =
	\frac{Q^2}{8} H(\as(\mu)) \int_0^\infty \rd b\, b\, J_0(b\, Q \sqrt{y}) S(Q,b)\,,
\label{eq:dEEC-res}
\eeq
where the large logarithmic corrections are exponentiated in the Sudakov form factor,
\beq
S(Q,b) = \exp\left\{-\int_{b_0^2/b^2}^{Q^2} \frac{\rd q^2}{q^2} 
	\left[A(\as(q^2)) \ln\frac{Q^2}{q^2} + B(\as(q^2))\right]\right\}\,.
\label{eq:dEEC-Sudakov}
\eeq
The Bessel function in \eqn{eq:dEEC-res} and $b_0 = 2 e^{-\gamma_{\rm E}}$ in
\eqn{eq:dEEC-Sudakov} have a kinematic origin. The functions $A$, $B$ and $H$ in 
\eqns{eq:dEEC-res}{eq:dEEC-Sudakov} are free of logarithmic corrections and can be 
computed as perturbative expansions in $\as$\footnote{Notice that our normalization 
conventions for $A^{(n)}$, $B^{(n)}$ and $H^{(n)}$, as well as $\beta_n$ in \eqn{eq:RGE} 
differ from \refr{deFlorian:2004mp}. We follow the conventions of \refr{Becher:2010tm}.},
\bal
A(\as) &= \sum_{n=1}^{\infty} \left(\frac{\as}{4\pi}\right)^n A^{(n)}\,,
\label{eq:A-func}
\\
B(\as) &= \sum_{n=1}^{\infty} \left(\frac{\as}{4\pi}\right)^n B^{(n)}\,,
\label{eq:B-func}
\\
H(\as) &= 1+ \sum_{n=1}^{\infty} \left(\frac{\as}{4\pi}\right)^n H^{(n)}\,.
\label{eq:H-func}
\eal

It is possible to perform the $q^2$ integration in \eqn{eq:dEEC-Sudakov} analytically and 
the Sudakov form factor can be written as
\beq
S(Q,b) = \exp\left[ 
	L g_1(a_{\rm S} \beta_0 L) 
	+ g_2(a_{\rm S} \beta_0 L) 
	+ a_{\rm S} g_3(a_{\rm S} \beta_0 L) + \ldots\right]\,,
\eeq
where $a_{\rm S} = \as(\mu)/(4\pi)$ and $L = \ln (Q^2 b^2/b_0^2)$ corresponds to $\ln y$ at 
large $b$ (the $y \ll 0$ limit corresponds to $Q b \gg 1$ through a Fourier transform). 
The $g_i$ functions read\footnote{We note that the form of the $g_i$ functions is not 
affected by our different choice of normalization as compared to \refr{deFlorian:2004mp}, 
hence our expressions agree with those in \refr{deFlorian:2004mp}.}
\bal
g_1(\lambda) &= 
	\frac{A^{(1)}}{\beta_0} \frac{\lambda+\ln(1-\lambda)}{\lambda} \,,
\label{eq:g1-func}
\\
g_2(\lambda) &= 
	\frac{B^{(1)}}{\beta_0} \ln(1-\lambda) -\frac{A^{(2)}}{\beta_0^2} 
	\left( \frac{\lambda}{1-\lambda} +\ln(1-\lambda)\right) - \frac{A^{(1)}}{\beta_0} 
	\left( \frac{\lambda}{1-\lambda} +\ln(1-\lambda)\right) \ln\xiRsq  
\nt \\& 
	+\frac{A^{(1)} \beta_1}{\beta_0^3} \left( \frac{1}{2} \ln^2(1-\lambda)+ 
	\frac{\ln(1-\lambda)}{1-\lambda} + \frac{\lambda}{1-\lambda}  \right) \,,
\label{eq:g2-func}
\\
g_3(\lambda) &= 
	-\frac{A^{(3)}}{2 \beta_0^2} \frac{\lambda^2}{(1-\lambda)^2}
	-\frac{B^{(2)}}{\beta_0} \frac{\lambda}{1-\lambda}
	+\frac{A^{(2)} \beta_1}{\beta_0^3} \left( \frac{\lambda (3\lambda-2)}{2(1-\lambda)^2} 
	- \frac{(1-2\lambda) \ln(1-\lambda)}{(1-\lambda)^2} \right) 
\nt \\ & 
	+ \frac{B^{(1)} \beta_1}{\beta_0^2} \left( \frac{\lambda}{1-\lambda} 
	+ \frac{\ln(1-\lambda)}{1-\lambda} \right) - \frac{A^{(1)}}{2} 
	\frac{\lambda^2}{(1-\lambda)^2}  \ln^2\xiRsq 
\nt \\ &
	- \ln\xiRsq \left[ B^{(1)} \frac{\lambda}{1-\lambda} 
	+ \frac{A^{(2)}}{\beta_0}\frac{\lambda^2}{(1-\lambda)^2} 
	+ A^{(1)} \frac{\beta_1}{\beta_0^2} \left( \frac{\lambda}{1-\lambda} 
	+ \frac{1-2\lambda}{(1-\lambda)^2} \ln(1-\lambda) \right) \right] 
\nt \\ & 
	+ A^{(1)} \bigg[ \frac{\beta_1^2}{2 \beta_0^4} \frac{1-2\lambda}{(1-\lambda)^2} 
	\ln^2(1-\lambda) + \ln(1-\lambda) \left(  \frac{\beta_0 \beta_2 -\beta_1^2}{\beta_0^4} 
	+ \frac{\beta_1^2}{\beta_0^4 (1-\lambda)}  \right)  
\nt\\ & 
	+ \frac{\lambda}{2 \beta_0^4 (1-\lambda)^2} ( \beta_0 \beta_2 (2-3\lambda)
	+ \beta_1^2 \lambda)\bigg] \,.
\label{eq:g3-func}
\eal

The functions $g_1$, $g_2$ and $g_3$ correspond to the LL, NLL and NNLL contributions.
The expansion coefficients $A^{(n)}$ and $B^{(n)}$ appearing in \eqnss{eq:g1-func}{eq:g3-func} 
above were obtained in \refr{deFlorian:2004mp} (see also \refrs{Becher:2010tm,Banfi:2014sua} 
for the complete NNLL $A^{(3)}$ coefficient). In our normalization conventions they read
\beq
\bsp
A^{(1)} &= 4\CF \,,
\\
A^{(2)} &= 
	\bigg[
	\CA \left(\frac{67}{9} - \frac{\pi^2}{3}\right) - \frac{20}{9}\Nf\TR 
	\bigg]\, A^{(1)} \,,
\\
A^{(3)} &= 
	\bigg[\CA^2 \bigg(\frac{245}{6} - \frac{134 \pi^2}{27} + \frac{11 \pi^4}{45}
	+ \frac{22}{3} \zeta_3\bigg) + \CF \Nf \TR \bigg(-\frac{55}{3} + 16 \zeta_3\bigg)
\\ &
	+\CA \Nf \TR \bigg(-\frac{418}{27} + \frac{40\pi^2}{27} - \frac{56}{3} \zeta_3\bigg) 
	- \frac{16}{27} \Nf^2 \TR^2 \bigg]\, A^{(1)} + 2 \beta_0 d^q_2\,,
\esp
\eeq
where
\beq
d^q_2 = \CA \CF \bigg(\frac{808}{27} - 28 \zeta_3\bigg) - \frac{224}{27} \CF \Nf \TR\,.
\eeq
For $B^{(1)}$ and $B^{(2)}$ we have
\beq
\bsp
B^{(1)} &= -6\CF \,,
\\
B^{(2)} &= -2\gamma_q^{(2)} - \CF \beta_0 \left(8 - \frac{10\pi^2}{3}  \right)\,,
\esp
\eeq
with
\beq
\gamma_q^{(2)} = \CF^2\left( \frac{3}{2} - 2\pi^2 + 24 \zeta_3 \right) 
	+ \CF \CA \left( \frac{17}{6} + \frac{22 \pi^2}{9} - 12 \zeta_3\right) 
	- \CF \Nf \TR\left( \frac{2}{3} + \frac{8 \pi^2}{9}\right)\,.
\eeq
Finally, $H^{(1)}$ reads
\beq
H^{(1)} = -\CF \left(11 + \frac{2\pi^2}{3}\right)\,,
\eeq
while the values of the higher loop coefficients $H^{(n)}$ ($n>1$) are currently unknown.

Notice that the $g_i$ functions are singular as $\lambda \to 1$. This singularity 
is related to the presence of the Landau pole in the QCD running coupling. Thus a 
prescription must be introduced to properly define the integral over $b$ in \eqn{eq:dEEC-res}. 
Here we follow the same procedure as in \refr{deFlorian:2004mp} and deform the contour of 
integration to the complex $b$-space as explained in \refrs{Laenen:2000de,Kulesza:2002rh,
Bozzi:2003jy}.

\begin{figure}[t]
\centering
\includegraphics[width=0.6\linewidth]{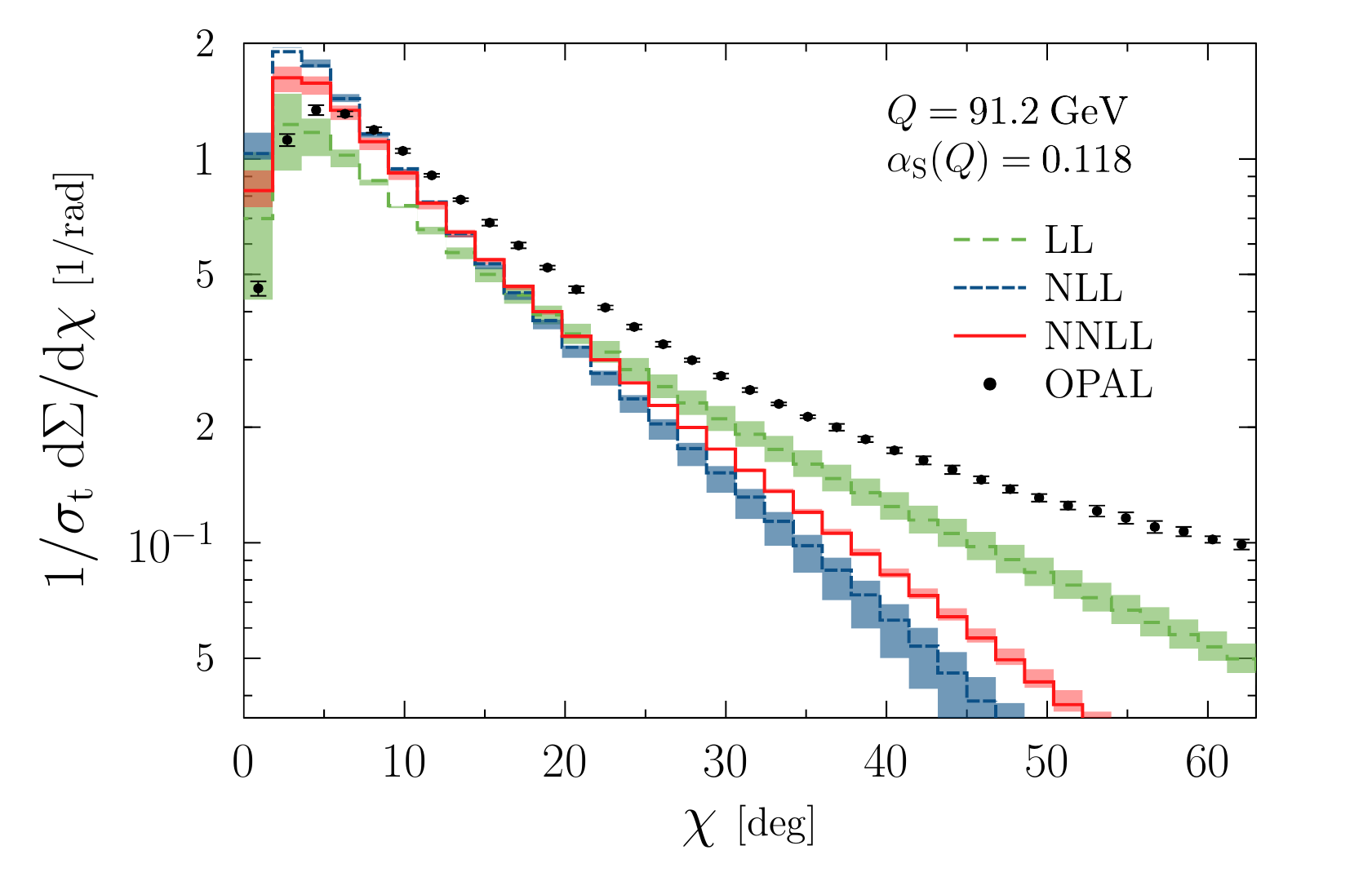}
\caption{Purely resummed predictions for EEC at LL, NLL and NNLL accuracy. 
The bands represent the effects of varying the renormalization scale by a factor of 
two around the default scale of $\mu = Q$. OPAL data is also shown.}
\label{fig:dEEC-res}
\end{figure}

In \fig{fig:dEEC-res} we present the resummed predictions for EEC up to NNLL accuracy 
together with OPAL data. Clearly, the resummed predictions are finite for $\chi \to 0$ 
and capture the general trends in the data in this limit. However, the purely resummed result 
badly underestimates the measured data already for moderate angles.
%


\section{Matching}
\label{sec:matching}

Fixed-order results are valid for moderate to large $y$ ($\as \ln^{2} y \ll 1)$, 
while resummed results apply to small $y$ ($y \ll 1$). In order to obtain predictions 
over a wide kinematical range\footnote{Notice that the description of the EEC distribution 
over the full angular range would require another resummation in the forward limit.} 
the two computations must be matched. A number of different matching procedures have 
been proposed in the literature (see for example \cite{Jones:2003yv} for a review), 
but conceptually they all involve adding the two computations and subtracting the 
doubly counted terms: 
\beq
\frac{1}{\tsigT} \frac{\rd \Sigma(\chi,\mu)}{\rd \cos \chi} =
	\bigg[\frac{1}{\tsigT} \frac{\rd \Sigma(\chi,\mu) }{\rd \cos \chi}\bigg]_{\mathrm{res.}}
	+ \bigg[\frac{1}{\tsigT} \frac{\rd \Sigma(\chi,\mu)}{\rd \cos \chi}\bigg]_{\mathrm{f.o.}}
	- \bigg\{
	\bigg[\frac{1}{\tsigT} \frac{\rd \Sigma(\chi,\mu)}{\rd \cos \chi}\bigg]_{\mathrm{res.}}
	\bigg\}\bigg|_{\mathrm{f.o.}}\,.
\label{eq:R-match-1}
\eeq
Here, the first term on the right hand side is the resummed result of \eqn{eq:dEEC-res}, 
the second term is the fixed-order prediction of \eqn{eq:dEEC-fo}, while the last term is 
obtained by expanding the resummed component to the same order in $\as$ as was 
used to compute the fixed-order result. Nevertheless, the subtraction of doubly counted terms 
alone is in general not sufficient to produce a physically sensible matched prediction. 
Indeed, in \eqn{eq:R-match-1} the resummed contribution on the right 
hand side is assumed to contain all logarithmically enhanced terms, hence the difference 
of the second and third terms,
\beq
\bigg[\frac{1}{\tsigT} \frac{\rd \Sigma(\chi,\mu)}{\rd \cos \chi}\bigg]_{\mathrm{f.o.}}
	- \bigg\{
	\bigg[\frac{1}{\tsigT} \frac{\rd \Sigma(\chi,\mu)}{\rd \cos \chi}\bigg]_{\mathrm{res.}}
	\bigg\}\bigg|_{\mathrm{f.o.}}\,,
\label{eq:R-match-1-diff}
\eeq
should be free of such contributions. However, unless the order of the logarithmic 
approximation is high enough to correctly reproduce the complete singular behavior 
of the fixed-order result as $\chi \to 0$, the difference in \eqn{eq:R-match-1-diff} will 
contain non-exponentiated subleading logarithmic terms which make the matched distribution 
divergent at small $\chi$. In contrast, the physical requirement is that the distribution 
should vanish at least as fast as a positive power of $\chi$. The matching procedure is thus 
in general more involved than a simple subtraction of the terms that have been doubly 
counted.

For EEC the NNLL approximation is sufficient to reproduce all singular terms in the 
LO and NLO fixed-order differential distributions. This is no longer true at NNLO 
accuracy. (Similarly the NLL approximation will only reproduce the complete singularity 
structure of the LO fixed-order result.) Hence \eqn{eq:R-match-1} may be used to define 
a matched result at NNLL+NLO accuracy but not at NNLL+NNLO accuracy.

In order to obtain a matched prediction at NNLL+NNLO accuracy which behaves physically 
for small $\chi$, the prescription of \eqn{eq:R-match-1} must be refined. 
This refinement has been worked out for EEC at NLL+NLO accuracy explicitly in 
\refr{Dokshitzer:1999sh} and corresponds to what is commonly referred to as `$R$ matching' 
for event shapes \cite{Catani:1992ua}. 
However, this matching scheme has the drawback that some matching coefficients must be 
extracted from the behavior of the fixed-order result around $\chi \to 0$. Since the 
fixed-order calculation is particularly challenging in this region, the matching coefficients 
can only be extracted with large numerical uncertainties. This issue becomes more severe 
as we go to higher orders in the fixed-order computation. Hence we will not 
develop $R$ matching for EEC at NNLL+NNLO. Nevertheless, the complete $R$ matching formula 
does simplify to just \eqn{eq:R-match-1} when applied at NNLL+NLO accuracy, thus we will 
refer to NNLL+NLO predictions obtained with \eqn{eq:R-match-1} as $R$ matched 
predictions below.

An alternative procedure for combining fixed-order and resummed results is log-$R$ 
matching \cite{Catani:1992ua}. An attractive feature of this scheme is that all matching 
coefficients can be extracted analytically from the resummed calculation. In the log-$R$ 
scheme one considers the cumulative event shape distribution, which we denote by the 
generic variable $R(y,\mu$) for some given event shape $y$:
\beq
R(y,\mu) = \frac{1}{\tsigT} \int_0^y \rd y'\, \frac{\dsig{}{}(y',\mu)}{\rd y'}\,.
\eeq
This quantity has the following fixed-order expansion:
\beq
\Big[R(y,\mu)\Big]_{\mathrm{f.o.}} = 1 + \frac{\as(\mu)}{2\pi} \bar{{\mathcal A}}(y,\mu)
	+ \left(\frac{\as(\mu)}{2\pi}\right)^2 \bar{{\mathcal B}}(y,\mu)
	+ \left(\frac{\as(\mu)}{2\pi}\right)^3 \bar{{\mathcal C}}(y,\mu)
	+ \Oa{4}\,,
\label{eq:R-fo}
\eeq
where the fixed-order coefficients $\bar{{\mathcal A}}$, $\bar{{\mathcal B}}$ and 
$\bar{{\mathcal C}}$ are obtained by integrating the corresponding differential 
distribution (e.g., \eqn{eq:dEEC-fo} for EEC) and using the constraint 
$R(y_{\mathrm{max}},\mu) = 1$ to all orders in $\as$ in order to fix the constants 
of integration. ($y_{\mathrm{max}}$ is the kinematically allowed maximum value of 
the variable $y$, for EEC $\chi_{\mathrm{max}} = 180\degree$.)

The specific formulas for log-$R$ matching in the literature \cite{Catani:1992ua} pertain 
to event shapes where the resummed prediction for the cumulative distribution can 
be written in a fully exponentiated form,
\beq
\Big[R(y,\mu)\Big]_{\mathrm{res.}} = (1 + C_1 \as + C_2 \as^2 + \ldots) 
	e^{L\, g_1(\as L) + g_2(\as L) + \as\, g_3(\as L) + \ldots} + {\mathcal O}(\as y)\,,
\label{eq:R-res}
\eeq
(with $L = \ln y$) where the $C_n$ and $g_n$ are known constants and functions. The 
$g_n$ functions can be expanded in powers of $\as$ and $L$ such that
\beq
g_n(\as L) = \sum_{i=1}^{\infty} G_{i,i+2-n} 
	\left(\frac{\as}{2\pi}\right)^i L^{i+2-n}\,.
\eeq
The log-$R$ matching scheme simply amounts to taking the logarithm of \eqn{eq:R-fo} 
and expanding it as a power series in $\as$ yielding:
\beq
\bsp
\ln \Big[R(y,\mu)\Big]_{\mathrm{f.o.}} &= 
	\frac{\as(\mu)}{2\pi} \bar{{\mathcal A}}(y,\mu)
	+ \left(\frac{\as(\mu)}{2\pi}\right)^2 
		\left[\bar{{\mathcal B}}(y,\mu) - \frac{1}{2} \bar{{\mathcal A}}(y,\mu)^2\right]
\\ &
	+ \left(\frac{\as(\mu)}{2\pi}\right)^3 
		\left[\bar{{\mathcal C}}(y,\mu) - \bar{{\mathcal A}}(y,\mu) \bar{{\mathcal B}}(y,\mu) 
		+ \frac{1}{3} \bar{{\mathcal A}}(y,\mu)^3\right] + \Oa{4}\,,
\esp
\label{eq:R-fo-exp}
\eeq
and similarly rewriting \eqn{eq:R-res} as:
\beq
\bsp
\ln \Big[R(y,\mu)\Big]_{\mathrm{res.}} &= 
	L\, g_1(\as L) + g_2(\as L) + \as\, g_3(\as L)
\\ &
	+ \as C_1 + \as^2 \left(C_2 - \frac{1}{2} C_1^2\right)
	+ \as^3 \left(C_3 - C_1 C_2 + \frac{1}{3} C_1^3\right) 
	+ \Oa{4}\,.
\esp
\label{eq:R-res-exp}
\eeq
Removing the terms up to $\Oa{3}$ from \eqn{eq:R-res-exp} and replacing them by the $\Oa{3}$ 
terms from \eqn{eq:R-fo-exp} yields the final expression for the log-$R$ matched prediction 
at NNLL+NNLO:
\bal
\ln R(y,\mu) &=  L g_1(\as L) + g_2(\as L) + \as g_3(\as L)
\label{eq:log-R-match-1}
\\&
	+ \frac{\as(\mu)}{2\pi} \left[\bar{\mathcal A}(y,\mu) - G_{11} L - G_{12} L^2\right]
\nt\\&
	+ \left(\frac{\as(\mu)}{2\pi}\right)^2 \left[\bar{\mathcal B}(y,\mu) 
		- \frac{1}{2} \bar{\mathcal A}^2(y,\mu) - G_{21} L - G_{22} L^2 - G_{23} L^3\right]
\nt\\&
	+ \left(\frac{\as(\mu)}{2\pi}\right)^3 \left[\bar{\mathcal C}(y,\mu) 
		- \bar{\mathcal A}(y,\mu) \bar{\mathcal B}(y,\mu) 
		+ \frac{1}{3} \bar{\mathcal A}^3(y,\mu) - G_{32} L^2 - G_{33} L^3 - G_{34} L^4\right]\,.
\nt
\eal
Notice that the constants $C_n$ do not enter \eqn{eq:log-R-match-1}, since they are removed 
from \eqn{eq:R-res-exp} and replaced by the fixed-order coefficients. Indeed, constant terms 
of the form $C_n \as^n$ must be factorized with respect to the form factor \cite{Catani:1992ua} 
and should not be exponentiated.

For the case of EEC, the straightforward application of \eqn{eq:log-R-match-1} faces two 
difficulties. First, the resummed expression is not directly in the form of \eqn{eq:R-res}. 
Second, EEC exhibits a particular problem because the fixed-order differential distribution  
diverges at both small and large $\chi$, so that the cumulative coefficients 
$\bar{\mathcal A}$, $\bar{\mathcal B}$ and $\bar{\mathcal C}$ 
cannot be reliably determined \cite{Acton:1993zh}. 

The latter complication can be conveniently solved by focusing on the following cumulative 
distribution:
\beq
\frac{1}{\tsigT} \widetilde{\Sigma}(\chi,\mu) \equiv 
	\frac{1}{\tsigT} \int_0^\chi \rd \chi'\, (1+\cos\chi') 
		\frac{\rd \Sigma(\chi',\mu)}{\rd \chi'} 
=
	\frac{1}{\tsigT} \int_0^{y(\chi)} \rd y'\, 2(1-y')  
		\frac{\rd \Sigma(y',\mu)}{\rd y'}\,,
\label{eq:EECcos-def}
\eeq
where we used the definition of $y$ in \eqn{eq:y-def} to obtain the second equality. 
Hence, $\widetilde{\Sigma}(\chi,\mu)/\tsigT$ is just a linear combination of the zeroth 
and first moments of the differential EEC distribution. It is straightforward to 
reconstruct the original differential EEC distribution from the quantity defined 
in \eqn{eq:EECcos-def}:
\beq
\frac{1}{\tsigT} \frac{\rd \Sigma(\chi,\mu)}{\rd\, \chi} = 
	\frac{1}{1 + \cos \chi} \frac{\rd}{\rd\, \chi} 
	\left[ \frac{1}{\tsigT} \widetilde{\Sigma}(\chi,\mu)\right]\,.
\eeq
In addition, $\widetilde{\Sigma}(\chi,\mu)$ has the following properties. First, the 
singularity in the forward region ($\chi \to \pi$ or $y\to 1$) of the differential EEC 
distribution present in the fixed-order perturbative predictions is regularized by the 
factor of $(1+\cos\chi)$ which goes to zero in this limit. Second, it is not difficult 
to show that in massless QCD the value of this cumulative distribution is unity when 
$\chi = 180\degree$, i.e., $\widetilde{\Sigma}(\chi_{\mathrm{max}},\mu)/\tsigT = 1$. 
These properties together ensure that the fixed-order cumulative coefficients for 
$\widetilde{\Sigma}$ (defined in \eqn{eq:R-fo} for a generic observable $R$) can be 
computed accurately by integrating the corresponding differential distribution and 
using $\widetilde{\Sigma}(\chi_{\mathrm{max}},\mu)/\tsigT = 1$ to all orders in $\as$ 
to fix the constants of integration.

Furthermore, using \eqn{eq:dEEC-res} in the definition of $\widetilde{\Sigma}$, 
\eqn{eq:EECcos-def}, we find that the integration over $y$ is straightforward to 
perform analytically and we obtain the following expression for the resummed prediction:
\beq
\bigg[\frac{1}{\tsigT} \widetilde{\Sigma}(\chi,\mu)\bigg]_{\mathrm{res.}} =
	\frac{H(\as(\mu))}{2} \int_0^\infty \rd b\, 
	\left[Q \sqrt{y} (1-y) J_1(b\, Q \sqrt{y}) 
		+ \frac{2 y}{b} J_2(b\, Q \sqrt{y}) \right] S(Q,b)\,,
\label{eq:EECcos-res}
\eeq
where the Sudakov form factor $S(Q,b)$ is unchanged and given in \eqn{eq:dEEC-Sudakov}.

Now let us turn to the issue of extending the definition of log-$R$ matching, 
\eqn{eq:log-R-match-1}, to the case of EEC. Although the resummed prediction 
for $\widetilde{\Sigma}$ is not directly in the form of \eqn{eq:R-res}, one can repeat the 
constructions in \eqnss{eq:R-fo-exp}{eq:log-R-match-1} to arrive at the analog 
of \eqn{eq:log-R-match-1} for $\widetilde{\Sigma}$. However, one must pay attention to 
the treatment of the non-logarithmically enhanced constant terms of the form $H^{(n)} \as^n$. 
These terms must not be exponentiated \cite{Catani:1992ua} and thus should not appear in 
the formula for the log-$R$ matched expression, just as the $C_n$ constants are absent 
in \eqn{eq:R-res-exp}. Hence, we find the following expression for EEC in the log-$R$ 
matching scheme\footnote{A similar expression was used in \refr{Acton:1993zh} to define 
a log-$R$ matched result for average jet multiplicity, see eq.~(20) of that paper.}
\beq
\bsp
\ln \bigg[\frac{1}{\tsigT} \widetilde{\Sigma}(\chi,\mu)\bigg] &= 
	\ln \bigg\{\frac{1}{H(\as(\mu))}
	\bigg[\frac{1}{\tsigT} \widetilde{\Sigma}(\chi,\mu)\bigg]_{\mathrm{res.}}\bigg\}
	+\frac{\as(\mu)}{2\pi} 
	\bigg[\bar{\mathcal A}(\chi,\mu) - \bar{\mathcal A}_{\mathrm{res.}}(\chi,\mu)\bigg]
\\&+
	\left(\frac{\as(\mu)}{2\pi}\right)^2 
	\left[\left(\bar{\mathcal B}(\chi,\mu) - \frac{1}{2} \bar{\mathcal A}^2(\chi,\mu)\right)
		-\left(\bar{\mathcal B}_{\mathrm{res.}}(\chi,\mu) 
			- \frac{1}{2} \bar{\mathcal A}_{\mathrm{res.}}^2(\chi,\mu)\right)
	\right]
\\&+
	\left(\frac{\as(\mu)}{2\pi}\right)^3
	\bigg[\left(\bar{\mathcal C}(\chi,\mu) - \bar{\mathcal A}(\chi,\mu) \bar{\mathcal B}(\chi,\mu) 
		+\frac{1}{3} \bar{\mathcal A}^3(\chi,\mu)\right)
\\&\qquad\qquad\qquad
		-\bigg(\bar{\mathcal C}_{\mathrm{res.}}(\chi,\mu) 
			- \bar{\mathcal A}_{\mathrm{res.}}(\chi,\mu) \bar{\mathcal B}_{\mathrm{res.}}(\chi,\mu) 
			+ \frac{1}{3} \bar{\mathcal A}_{\mathrm{res.}}^3(\chi,\mu)\bigg)
	\bigg]\,,
\esp
\label{eq:EECcos-log-R-match}
\eeq
where $\bar{\mathcal A}_{\mathrm{res.}}$, $\bar{\mathcal B}_{\mathrm{res.}}$ 
and $\bar{\mathcal C}_{\mathrm{res.}}$ are the coefficients obtained by expanding 
the resummed component in the curly brackets above in a power series in $\as$:
\beq
\bsp
\frac{1}{H(\as(\mu))}
	\bigg[\frac{1}{\tsigT} \widetilde{\Sigma}(\chi,\mu)\bigg]_{\mathrm{res.}} &=
	1
	+
	\frac{\as(\mu)}{2\pi} \bar{\mathcal A}_{\mathrm{res.}}(\chi,\mu) 
	+
	\left(\frac{\as(\mu)}{2\pi}\right)^2 \bar{\mathcal B}_{\mathrm{res.}}(\chi,\mu) 
\\&\qquad
	+
	\left(\frac{\as(\mu)}{2\pi}\right)^3 \bar{\mathcal C}_{\mathrm{res.}}(\chi,\mu)
	+
	\Oa{4}\,.
\label{eq:EECcos-res-exp}
\esp
\eeq
The explicit expressions for these coefficients are somewhat long and we present them 
in \appx{appx:RES-EXP}.

\Eqn{eq:EECcos-log-R-match} is our final result for the log-$R$ matched prediction for 
$\widetilde{\Sigma}$ at NNLL+NNLO accuracy. The log-$R$ matched prediction at NNLL+NLO 
accuracy is obtained simply by dropping the $\Oa{3}$ term in \eqn{eq:EECcos-log-R-match}. 
We emphasize that the quantities $H^{(n)}$ do not appear in \eqn{eq:EECcos-log-R-match} 
at all. In the log-$R$ matching scheme these terms, as well as subdominant logarithmic 
contributions are all implicit in the unsubtracted parts of the fixed-order coefficients 
$\bar{\mathcal A}$, $\bar{\mathcal B}$ and $\bar{\mathcal C}$  \cite{Catani:1992ua}. 
Hence the log-$R$ matched prediction can be computed without the explicit knowledge of 
$H^{(n)}$.


\section{Phenomenological results}
\label{sec:pheno}

In the following we investigate the numerical impact of the NNLO corrections 
and show quantitative predictions for the differential EEC distribution at 
NNLL+NLO and NNLL+NNLO accuracy.

\begin{figure}[t]
\centering
\includegraphics[width=0.6\linewidth]{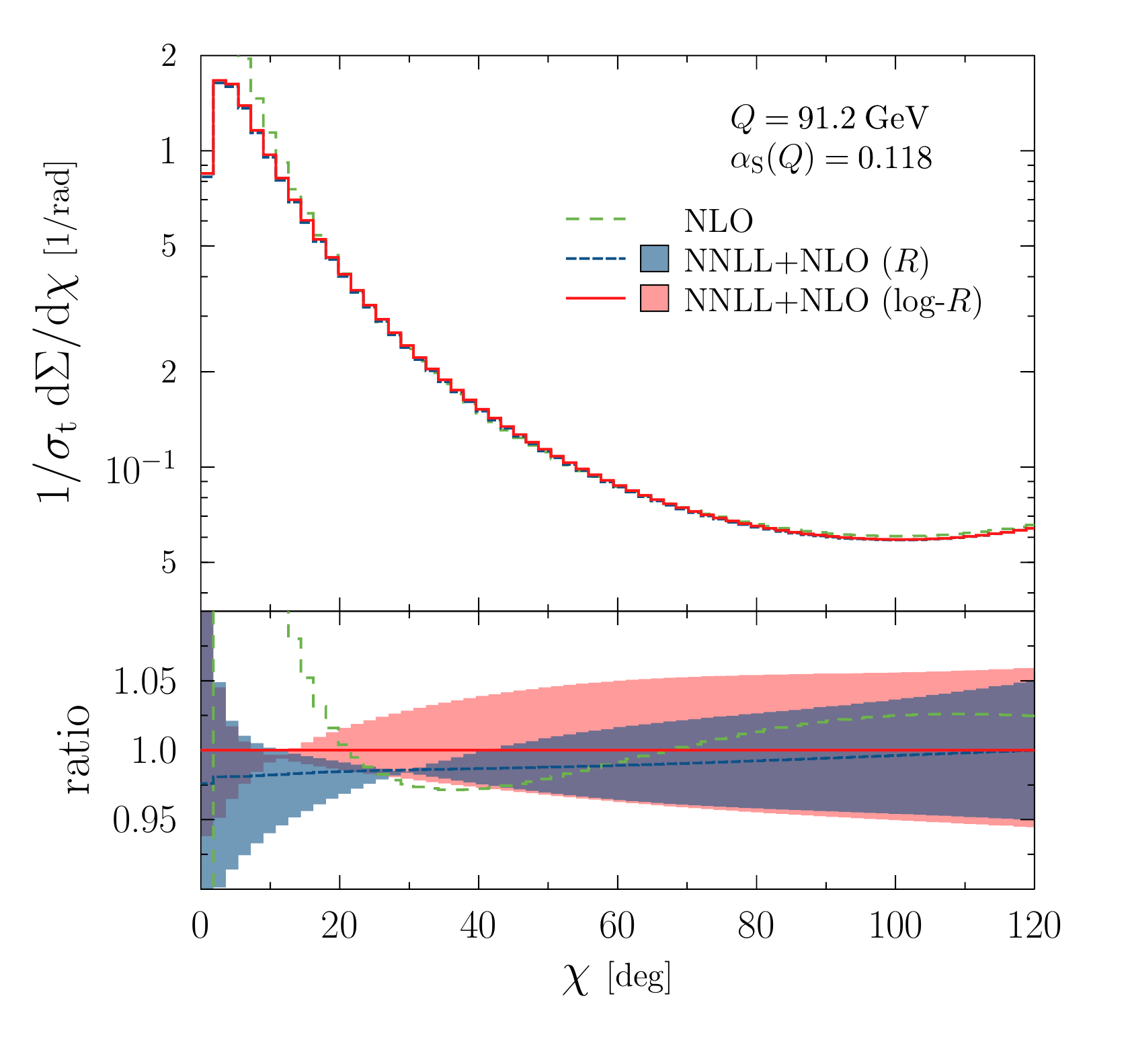}
\caption{NNLL+NLO matched predictions for EEC using $R$ and log-$R$ matching. The bottom 
panel shows the ratio of the fixed-order NLO and $R$ matched predictions to the log-$R$ 
matched result. The bands represent renormalization scale variation of the matched 
predictions in the range $\mu \in [Q/2,2Q]$ with two-loop running of $\as$.}
\label{fig:NNLO+NLL-R}
\end{figure}
We start by presenting the NNLL+NLO predictions. As discussed above, at this 
accuracy both R matching (\eqn{eq:R-match-1}) as well as log-$R$ matching 
(\eqn{eq:EECcos-log-R-match}) may be used to define physically sensible matched 
predictions. The results obtained with both the $R$ matching and log-$R$ matching 
schemes are shown in \fig{fig:NNLO+NLL-R}, together with the fixed-order NLO result. 
Throughout we set the center-of-mass energy to the $Z$-boson mass, 
$Q=M_Z=91.2$~GeV while the strong coupling is fixed to $\as(M_Z) = 0.118$. The 
fixed-order prediction is seen to diverge to $-\infty$ as $\chi \to 0$. On the other 
hand, the matched results remain well-behaved in both matching schemes down to very 
small values of $\chi$.
In the bottom panel of \fig{fig:NNLO+NLL-R} we show the ratio of the fixed-order NLO and 
the $R$ matched predictions to the log-$R$ matched result. The bands represent the effect 
of varying the renormalization scale by a factor of two around its central value of $\mu=Q$, 
using two-loop running for the strong coupling. We see that the two matching schemes give 
very similar results with the relative difference of the $R$ matched prediction to the 
log-$R$ matched prediction changing from about $-2\%$ at $\chi \sim 0\degree$ to $0\%$ for 
$\chi \sim 120\degree$. Around $\chi \sim 180\degree$, the relative difference is about 
$+0.5\%$.

\begin{figure}[t]
\centering
\includegraphics[width=0.6\linewidth]{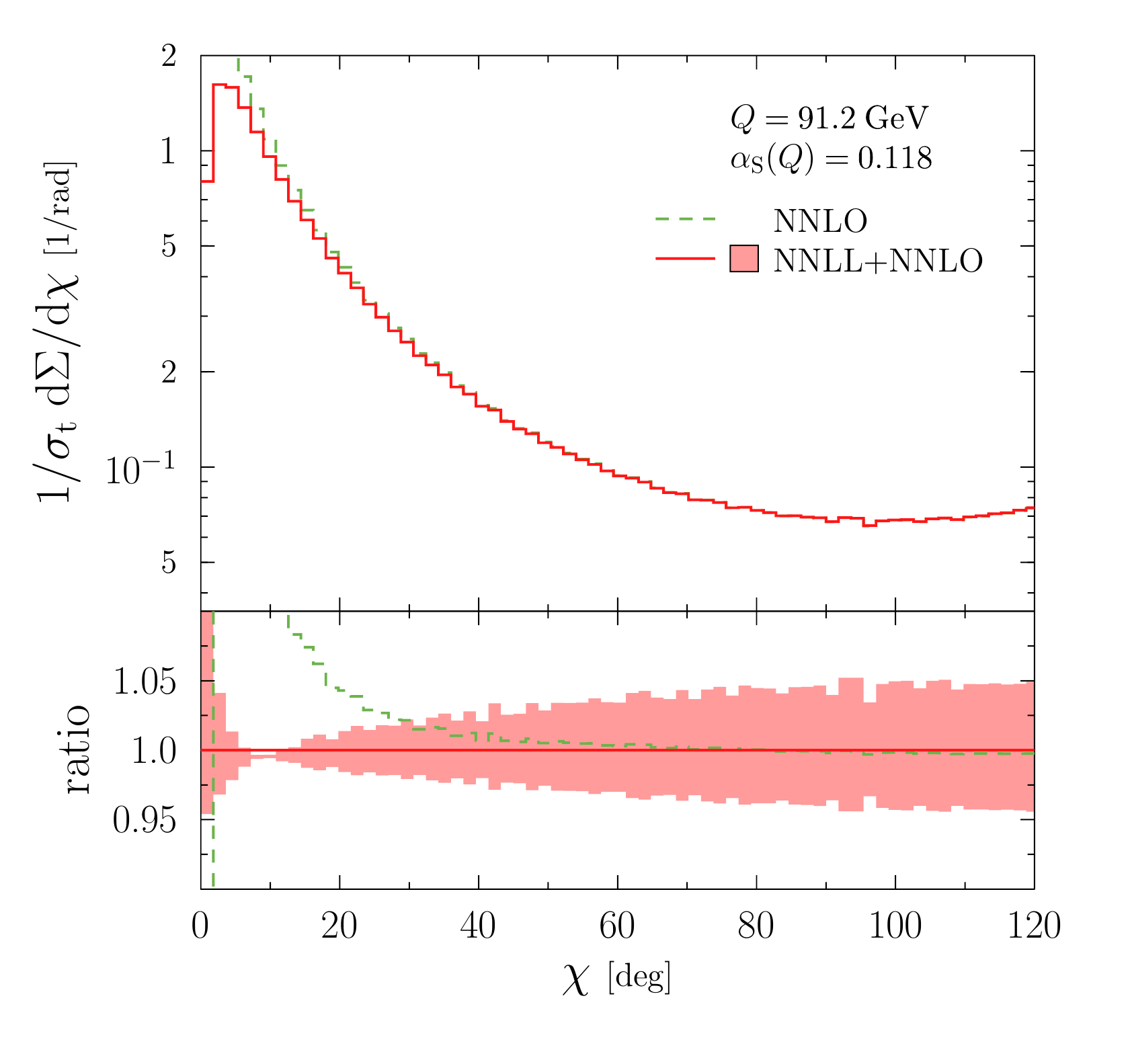}
\caption{NNLL+NNLO matched prediction for EEC. The bottom panel shows the ratio of the 
fixed-order NNLO prediction to the matched result. The band represents renormalization 
scale variation of the matched prediction in the range $\mu \in [Q/2,2Q]$ with three-loop 
running of $\as$.}
\label{fig:NNLO+NNLL-log-R}
\end{figure}
Next, we include the NNLO corrections and present our NNLL+NNLO results. At this accuracy 
only the log-$R$ matching scheme gives a prediction which behaves physically at small $\chi$, 
and this prediction is shown in \fig{fig:NNLO+NNLL-log-R}. Here too, the center-of-mass 
energy was set to $Q=91.2$~GeV and we used $\as(M_Z) = 0.118$. The fixed-order prediction at 
NNLO accuracy diverges to $+\infty$ as $\chi \to 0$. (The divergence to $+\infty$ is not 
visible on the plot where the NNLO result seems to diverge to $-\infty$.) As previously, 
the matched prediction is well-behaved down to very small values of $\chi$. The bottom 
panel shows the ratio of the fixed-order prediction to the matched result. The band again 
represents the effect of varying the renormalization scale by a factor of two in either 
direction around its central value of $\mu = Q$. In this case, three-loop running of the 
strong coupling is used. 

\begin{figure}[t]
\centering
\includegraphics[width=0.6\linewidth]{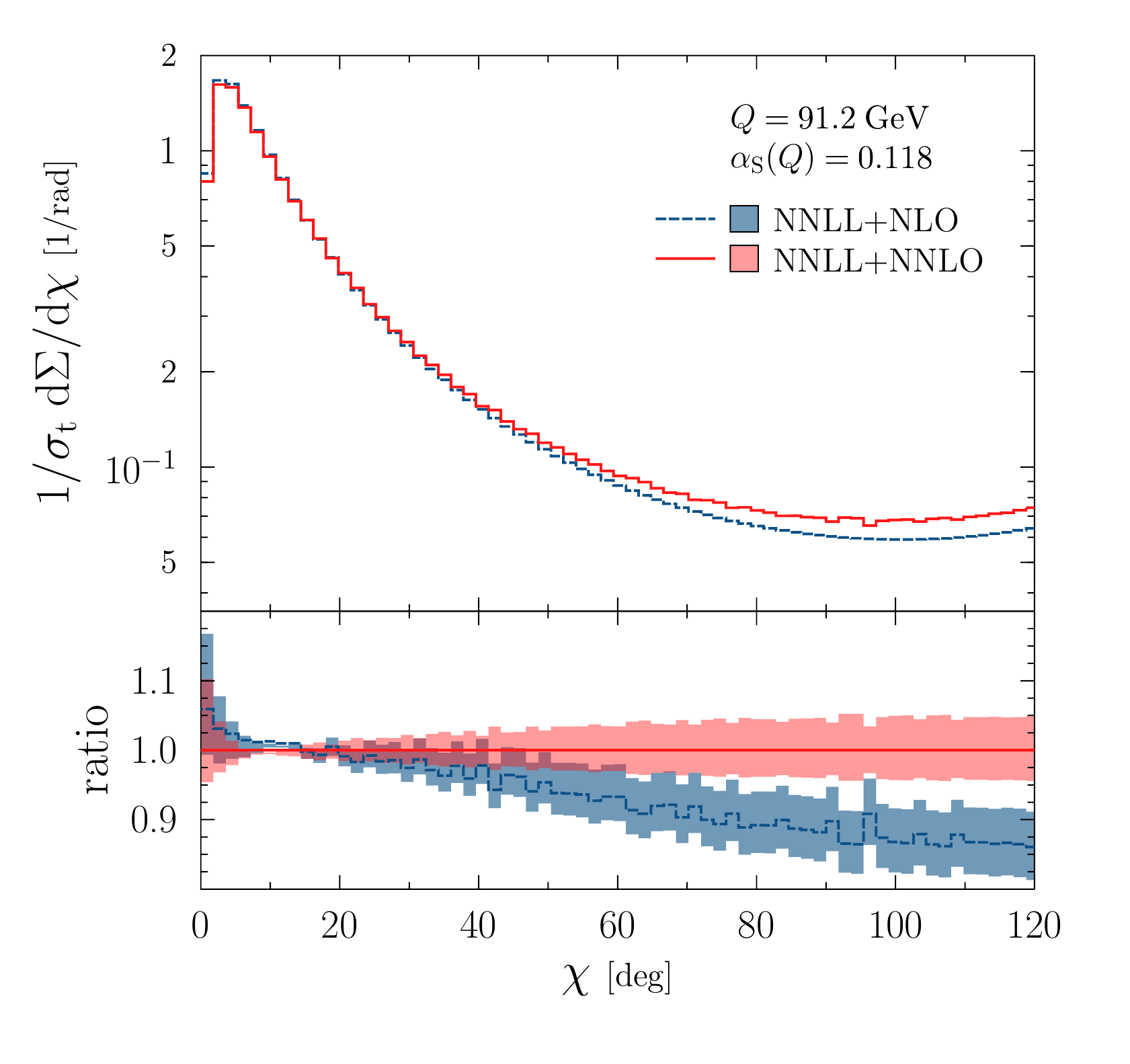}
\caption{Matched predictions for EEC at NNLL+NLO and NNLL+NNLO accuracy in the log-$R$ 
matching scheme. The bottom panel shows the ratio of the NNLL+NLO result to the NNLL+NNLO 
prediction. The bands represent renormalization scale variation in the range 
$\mu \in [Q/2,2Q]$.}
\label{fig:NNLO+NNLL-vs-NLO+NNLL-log-R}
\end{figure}
To better appreciate the impact of the NNLO corrections on the matched prediction, 
in \fig{fig:NNLO+NNLL-vs-NLO+NNLL-log-R} we compare the NNLL+NLO and NNLL+NNLO results 
obtained in the log-$R$ matching scheme. The lower panel shows the ratio of the NNLL+NLO 
prediction to the NNLL+NNLO one. We see that the inclusion of the NNLO corrections slightly 
lowers the prediction in and below the peak region by from $-5\%$ to $-2\%$, while the 
prediction is enhanced for medium and high values of $\chi$. This enhancement goes from 
around $+7\%$ at $\chi = 60\degree$ to around $+14\%$ at $\chi = 120\degree$ and up 
to $+20\%$ to $+25\%$ for values of $\chi$ near $180\degree$. 
Hence the inclusion of NNLO corrections has a sizable impact on the shape of the distribution.

Next, we compare our predictions to precise OPAL \cite{Acton:1993zh} and SLD \cite{Abe:1994mf} 
data. In particular, we perform a fit of our most accurate NNLL+NNLO prediction 
to the experimental data with the strong coupling $\as$ as a free parameter. We use a $\chi^2$ 
analysis for the fitting procedure. In general, both statistical and systematic errors are 
correlated between bins, but unfortunately the experimental publications provide practically 
no information on the correlations. Therefore, we simply add statistical and systematic 
uncertainties in quadrature and treat them as uncorrelated between all data points. 
In order to quantify the impact of the NNLO corrections, we perform the same fit also for 
the NNLL+NLO prediction computed in both the $R$ and the log-$R$ matching schemes. 
Since we cannot take into account correlations properly, we do not aim to produce the most 
accurate extraction of $\as(M_Z)$ here, but rather to assess the impact of the NNLO corrections 
on the extraction.

In a first attempt, we neglect hadronization corrections, however, we come back to this 
point below. Obviously, the results obtained in this way must be interpreted with care.

In order to ease the comparison of our results to previous work, we choose the fit 
ranges of \refr{deFlorian:2004mp} for our analyses. In the first case we include data 
in the range $0\degree < \chi < 63\degree$ where the effects of resummation are rather 
pronounced. However, given that the low $\chi$ region is particularly sensitive to 
non-perturbative corrections, we also investigate the range $15\degree < \chi < 63\degree$ 
where the lower cut is expected to mitigate the effects of these contributions.   
Finally, we perform fits including data in the $15\degree < \chi < 120\degree$ interval.
Since for large $\chi$ the matched prediction is controlled by the fixed-order result, 
we expect the effects of the NNLO correction to be most prominent here. The upper limit 
is chosen to cut the forward region where another resummation would be required.
\begin{table}
\renewcommand*{\arraystretch}{1.5}
\centering
  \begin{tabular}{ccccccc}
    \toprule
    \multirow{2}{*}{Fit range} &
      \multicolumn{2}{c}{NNLL+NLO ($R$)} &
      \multicolumn{2}{c}{NNLL+NLO (log-$R$)} &
      \multicolumn{2}{c}{NNLL+NNLO (log-$R$)} \\
      & {$\as(M_Z)$} & {$\chi^2$/d.o.f.} & {$\as(M_Z)$} & {$\chi^2$/d.o.f.} & {$\as(M_Z)$} & {$\chi^2$/d.o.f.} \\
      \midrule
    $0\degree < \chi < 63\degree$ & 
    		$0.133 \pm 0.001$ & $1.96$ & 
		$0.131 \pm 0.003$ & $1.21$ & 
		$0.129 \pm 0.003$ & $4.13$ \\
    $15\degree < \chi < 63\degree$ & 
    		$0.132 \pm 0.001$ & $0.59$ & 
		$0.131 \pm 0.003$ & $0.54$ & 
		$0.128 \pm 0.003$ & $1.58$ \\
    $15\degree < \chi < 120\degree$ & 
    		$0.135 \pm 0.002$ & $3.96$ & 
		$0.134 \pm 0.004$ & $5.12$ & 
		$0.127 \pm 0.003$ & $1.12$ \\
    \bottomrule
  \end{tabular}
\caption{Results of the fits of the matched predictions at NNLL+NLO and NNLL+NNLO accuracy 
to OPAL and SLD data. The number of degrees of freedom of the fits are 
$\mbox{d.o.f.}=50$ for $0\degree < \chi < 63\degree$, 
$\mbox{d.o.f.}=38$ for $15\degree < \chi < 63\degree$ and 
$\mbox{d.o.f.}=86$ for $15\degree < \chi < 120\degree$.}
\label{tab:PT-fits}
\end{table}

We collect the results of these fits in \tab{tab:PT-fits} where we show the best fit value 
of $\as$ and the $\chi^2/\mathrm{d.o.f.}$ for each fit. The quoted uncertainty on the 
extracted value of $\as$ is obtained by adding the fit uncertainty and the theoretical 
uncertainty from missing higher-order contributions in quadrature. We assess the latter 
by repeating the fit with several values of $\mu$ in the range $Q/2 \le \mu \le 2Q$ and 
taking the envelope of the obtained results. This theoretical contribution is dominant in 
the total uncertainty.

In the first case, when we include data in the $0\degree < \chi < 63\degree$ interval, 
we observe that the quality of the fit is actually better for the NNLL+NLO predictions 
than the theoretically most accurate NNLL+NNLO prediction, with the latter fit being 
rather poor as evidenced by the high value of $\chi^2/\mathrm{d.o.f.} = 206.4/50 = 4.13$. 
The extracted values of $\as(M_Z)$ are generally quite high compared to the world average 
$\as(M_Z) = 0.1181 \pm 0.0011$ \cite{Bethke:2017uli}, 
in the range $\as(M_Z) = 0.129$ -- $0.133$, with the fit using the NNLL+NNLO prediction 
giving the smallest value. However, as mentioned above, non-perturbative corrections are 
the largest in the low-$\chi$ region, hence the results of fits using purely perturbative 
predictions should be interpreted with great care, especially in this fit range.

Indeed, repeating the fit in the $15\degree < \chi < 63\degree$ interval, we observe that 
the $\chi^2/\mathrm{d.o.f.}$ decreases in each case and the change is particularly 
significant for the fit using the NNLL+NNLO prediction, where we find a much more 
reasonable value of $\chi^2/\mathrm{d.o.f.} = 60.1/38 = 1.58$. On the other hand, the 
extracted values of $\as(M_Z)$ are rather insensitive to this lower cut on the data and 
are still high compared to the world average.

Finally, we perform the fits including data in the $15\degree < \chi < 120\degree$ 
interval. The quality of the fits based on NNLL+NLO predictions deteriorates quite 
drastically as evidenced by the rather high values of $\chi^2/\mathrm{d.o.f} = 340.3/86 
= 3.96$ for the $R$ matched prediction and $\chi^2/\mathrm{d.o.f.} = 440.1/86 = 5.12$ 
for the log-$R$ matched one. At the same time the extracted values of $\as(M_Z)$ become 
even higher with $\as(M_Z) = 0.134$ -- $0.135$. However, the inclusion of NNLO correction 
drastically improves the quality of the fit and we obtain $\chi^2/\mathrm{d.o.f.} = 95.9/86 
= 1.12$. 
The extracted value of $\as(M_Z)$ also decreases somewhat and we find 
$\as(M_Z) = 0.127 \pm 0.003$ for the best fit value. 

Our extracted values of $\as(M_Z)$ based on the NNLL+NLO predictions using 
$R$ matching are quite close to the values obtained in \refr{deFlorian:2004mp} for all three 
fit ranges, although our results are marginally higher. We have checked that these differences 
are due to the fact that the determinations in \refr{deFlorian:2004mp} used the incomplete 
$A^{(3)}$ NNLL resummation coefficient.

Overall, we observe that the inclusion of the fixed-order NNLO corrections reduces the 
extracted value of $\as(M_Z)$. This reduction is about $-2\%$ to $-3\%$ when data in the 
range $0\degree < \chi < 63\degree$ are taken into account, about $-2\%$ to $-4\%$ for the 
range $15\degree < \chi < 63\degree$ and between $-5\%$ to $-7\%$ when 
$15\degree < \chi < 120\degree$, depending on the matching prescription used for the NNLL+NLO 
prediction. Hence, these corrections must be included in a precise determination of 
$\as$ using EEC.

In our analysis so far, we have neglected hadronization corrections. However, 
non-perturbative contributions are expected to be relevant, especially at small 
angles \cite{Basham:1978bw,Collins:1981uk,Collins:1981va,Collins:1985xx,Fiore:1992sa}, 
and indeed the OPAL analysis of \refr{Acton:1993zh} found hadron-parton correction 
factors from around 1.5 for very small $\chi$ to around 0.9 for large $\chi$\footnote{In \refr{Acton:1993zh} only the hadron level data is given in a tabulated form with uncertainties, 
while the parton level data appears only in plots. This is nevertheless sufficient to 
assess the magnitude of the hadron-parton correction factors even without the original 
parton level data.}. Hence 
it is important to account for these non-perturbative contributions. As already mentioned, 
these can be determined either by extracting them from data by comparison 
to Monte Carlo predictions, or by performing analytic model calculations. Here, we follow 
the latter option and use the non-perturbative model of \refr{Dokshitzer:1999sh} to 
describe the hadronization contributions. Thus we multiply the Sudakov form factor of 
\eqn{eq:dEEC-Sudakov} with a correction of the form
\beq
S_{\mathrm{NP}} = e^{-\frac{1}{2} a_1 b^2} (1 - 2 a_2 b)\,,
\label{eq:dEEC-Sudakov-NP}
\eeq
and treat $a_1$ and $a_2$ as free parameters of the non-perturbative model to be fitted 
from data.

We have performed a three-parameter fit including data in the $0\degree < \chi < 63\degree$ 
range using our NNLL+NNLO prediction, as well as the predictions obtained at NNLL+NLO 
with both $R$ matching and log-$R$ matching. In the $R$ matching scheme at NNLL+NLO accuracy, 
we extract the following parameters:
\beq
\mbox{NNLL+NLO ($R$):}\quad
\as(M_Z) = 0.134^{+0.001}_{-0.009}\,,
\quad
a_1 = 1.55^{+4.26}_{-1.54}\; \mathrm{GeV}^2\,,
\quad
a_2 = -0.13^{+0.50}_{-0.05}\; \mathrm{GeV}\,,
\label{eq:NNLL+NLO-R-3p-fit}
\eeq
with $\chi^2/\mathrm{d.o.f.} = 38.7/48 = 0.81$. All uncertainties are again obtained 
by adding the fit uncertainties and the theoretical uncertainties in quadrature. 
The theoretical uncertainties are assessed by varying the renormalization scale $\mu$ 
between $Q/2$ and $2Q$ and repeating the fit. The total uncertainties are mostly dominated 
by the theoretical uncertainty with the exception of the upper limit of strong coupling. 
In this case, we find that the maximal best fit value of $\as$ is obtained for $\mu \simeq Q$, 
hence the upper limit is controlled by the fit uncertainty. We also report the correlation 
matrix of the fit for the central values:
\beq
\mbox{NNLL+NLO ($R$):}\quad
\mathrm{corr}(\as, a_1, a_2) = 
\begin{pmatrix} 
1 & 0.04 & -0.70 \\
0.04 & 1 & -0.03 \\
-0.70 & -0.03 & 1
\end{pmatrix}\,.
\label{eq:NNLL+NLO-R-corr}
\eeq
Evidently the strong coupling $\as$ is highly anti-correlated with the non-perturbative 
parameter $a_2$.

The analysis of \refr{deFlorian:2004mp} performed on the same data gave 
$|a_2| \lesssim 0.002\; \mathrm{GeV}$, a very small value compatible with $a_2 = 0$. 
After fixing the parameter $a_2$ to zero, a two-parameter fit to the strong coupling and 
the remaining non-perturbative parameter $a_1$ produced the best fit values of 
$\as(M_Z) = 0.130^{+0.002}_{-0.004}$ and $a_1 = 1.5^{+3.2}_{-0.5}\; \mathrm{GeV}^2$ 
with $\chi^2/\mathrm{d.o.f.} = 0.99$. Our results in \eqn{eq:NNLL+NLO-R-3p-fit} are 
compatible with these values within uncertainties. We have nevertheless verified that 
the source of the discrepancy between the two extractions is, again, due to the fact that 
\refr{deFlorian:2004mp} used the incomplete $A^{(3)}$ NNLL resummation coefficient.

Turning to the log-$R$ matching scheme at NNLL+NLO accuracy, we obtain the results:
\beq
\mbox{NNLL+NLO (log-$R$):}\quad
\as(M_Z) = 0.128^{+0.002}_{-0.006}\,,
\quad
a_1 = 1.17^{+1.46}_{-0.29}\; \mathrm{GeV}^2\,,
\quad
a_2 = 0.13^{+0.14}_{-0.09}\; \mathrm{GeV}\,,
\label{eq:NNLL+NLO-log-R-3p-fit}
\eeq
and we find $\chi^2/\mathrm{d.o.f.} = 40.8/48 = 0.85$, with the correlation matrix for 
the central values
\beq
\mbox{NNLL+NLO (log-$R$):}\quad
\mathrm{corr}(\as, a_1, a_2) = 
\begin{pmatrix} 
1 & -0.17 & -0.98 \\
-0.17 & 1 & 0.08 \\
-0.98 & 0.08 & 1
\end{pmatrix}\,.
\label{eq:NNLL+NLO-log-R-corr}
\eeq
The strong coupling $\as$ and the $a_2$ non-perturbative parameter is even more strongly 
anti-correlated than in the $R$ matching scheme. As before, the uncertainties in 
\eqn{eq:NNLL+NLO-log-R-3p-fit} include the fit and theoretical uncertainties added in 
quadrature. We observe that the quality of the fits as measured by $\chi^2/\mathrm{d.o.f.}$ 
is very similar in the two matching schemes and the fit results are compatible between the 
two schemes within uncertainties. The extracted value of the strong coupling is reduced by 
about $-5\%$ in the log-$R$ scheme compared to the $R$ scheme, however, it remains high 
compared to the world average in both schemes.

\begin{figure}[t]
\centering
\includegraphics[width=0.6\linewidth]{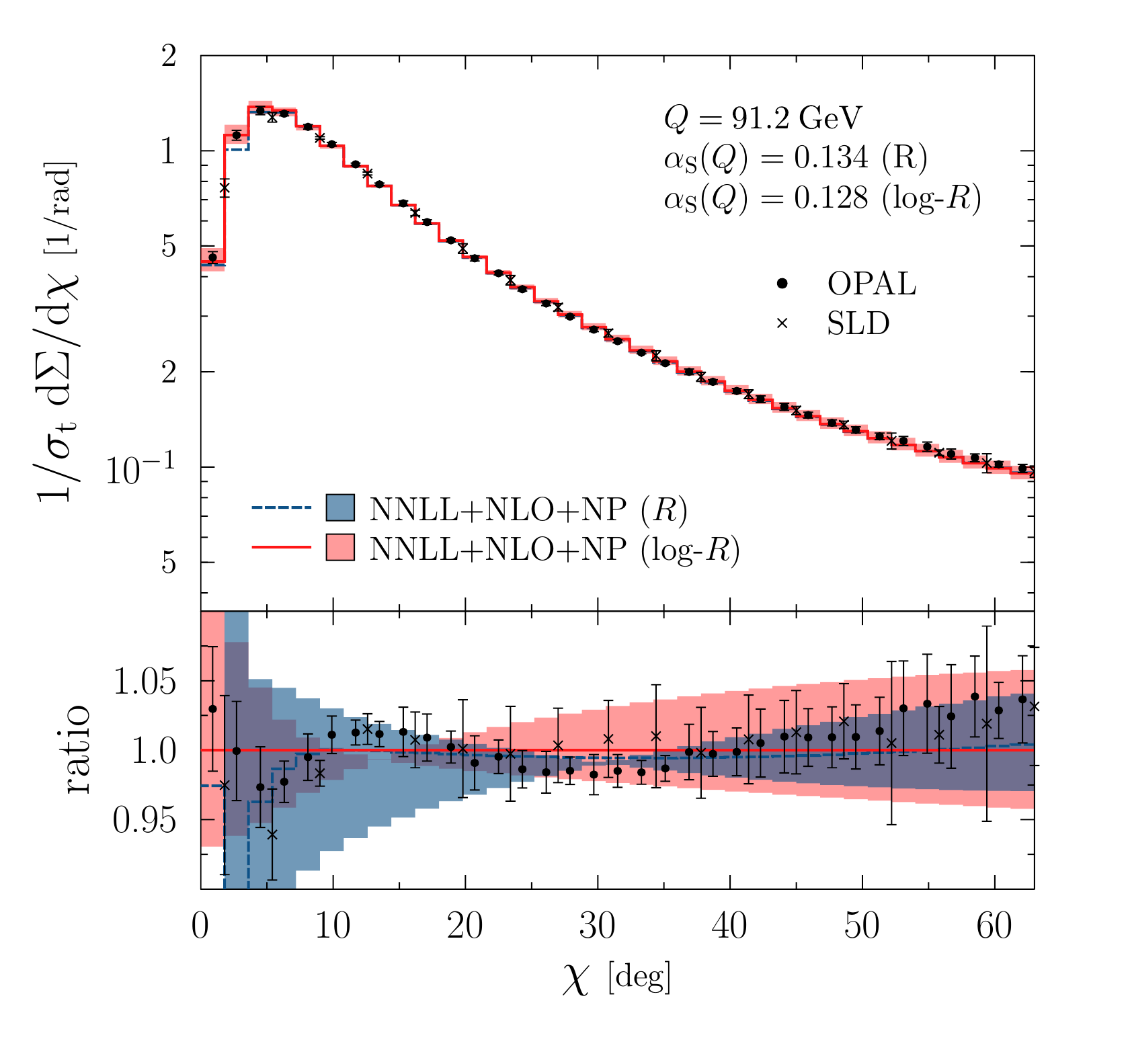}
\caption{NNLL+NLO matched predictions for EEC in the $R$ and log-$R$ matching schemes. 
The analytic model of \eqn{eq:dEEC-Sudakov-NP} is used to account for hadronization 
corrections. The bottom panel shows the ratio of the data and the $R$ matched prediction 
to the log-$R$ matched result. The bands represent the effect of varying the renormalization 
scale in the range $\mu \in [Q/2,2Q]$ with two-loop running of $\as$.}
\label{fig:NNLL+NLO-3p-fit}
\end{figure}
We present the comparison of the best fit NNLL+NLO predictions in the $R$ and log-$R$ matching 
schemes to the data in \fig{fig:NNLL+NLO-3p-fit}. The figure shows a nice overall agreement 
between the predictions and experiment and it is clear that the calculations can reproduce the  
measurements up to the smallest measured values of $\chi$. Nevertheless, we observe a small 
but systematic deviation of the prediction from data in the region of medium $\chi$ (from 
about $\chi \gtrsim 30\degree$) and it is clear that the shape of the measured distribution 
is not fully reproduced. The bottom panel shows the ratio of the data and the $R$ matched 
prediction to the log-$R$ matched result, with the bands representing scale uncertainty. 

Finally, we investigate the impact of NNLO corrections and repeat the three-parameter fit 
in the same range of $0\degree < \chi < 63\degree$, but using our most accurate NNLL+NNLO 
theoretical prediction. The best fit corresponds to $\chi^2/\mathrm{d.o.f.} = 56.7/48 
= 1.18$ and we extract the following parameter values:
\beq
\mbox{NNLL+NNLO (log-$R$):}\quad
\as(M_Z) = 0.121^{+0.001}_{-0.003}\,,
\quad
a_1 = 2.47^{+0.48}_{-2.38}\; \mathrm{GeV}^2\,,
\quad
a_2 = 0.31^{+0.27}_{-0.05}\; \mathrm{GeV}\,.
\eeq
Once more, the uncertainties shown include the fit uncertainties and theoretical 
uncertainties added in quadrature. The correlation matrix of the fit for the central 
values again shows that $\as$ and $a_2$ are very strongly anti-correlated:
\beq
\mbox{NNLL+NNLO (log-$R$):}\quad
\mathrm{corr}(\as, a_1, a_2) = 
\begin{pmatrix} 
1 & 0.05 & -0.97 \\
0.05 & 1 & -0.07 \\
-0.97 & -0.07 & 1
\end{pmatrix}\,.
\eeq
We see that the quality of the fit improves drastically compared to the purely perturbative 
fit reported in \tab{tab:PT-fits}. Moreover, the extracted value of $\as(M_Z)$ is sizably 
reduced compared to the fits based on NNLL+NLO predictions and is indeed compatible with 
the world average within uncertainties. 

\begin{figure}[t]
\centering
\includegraphics[width=0.6\linewidth]{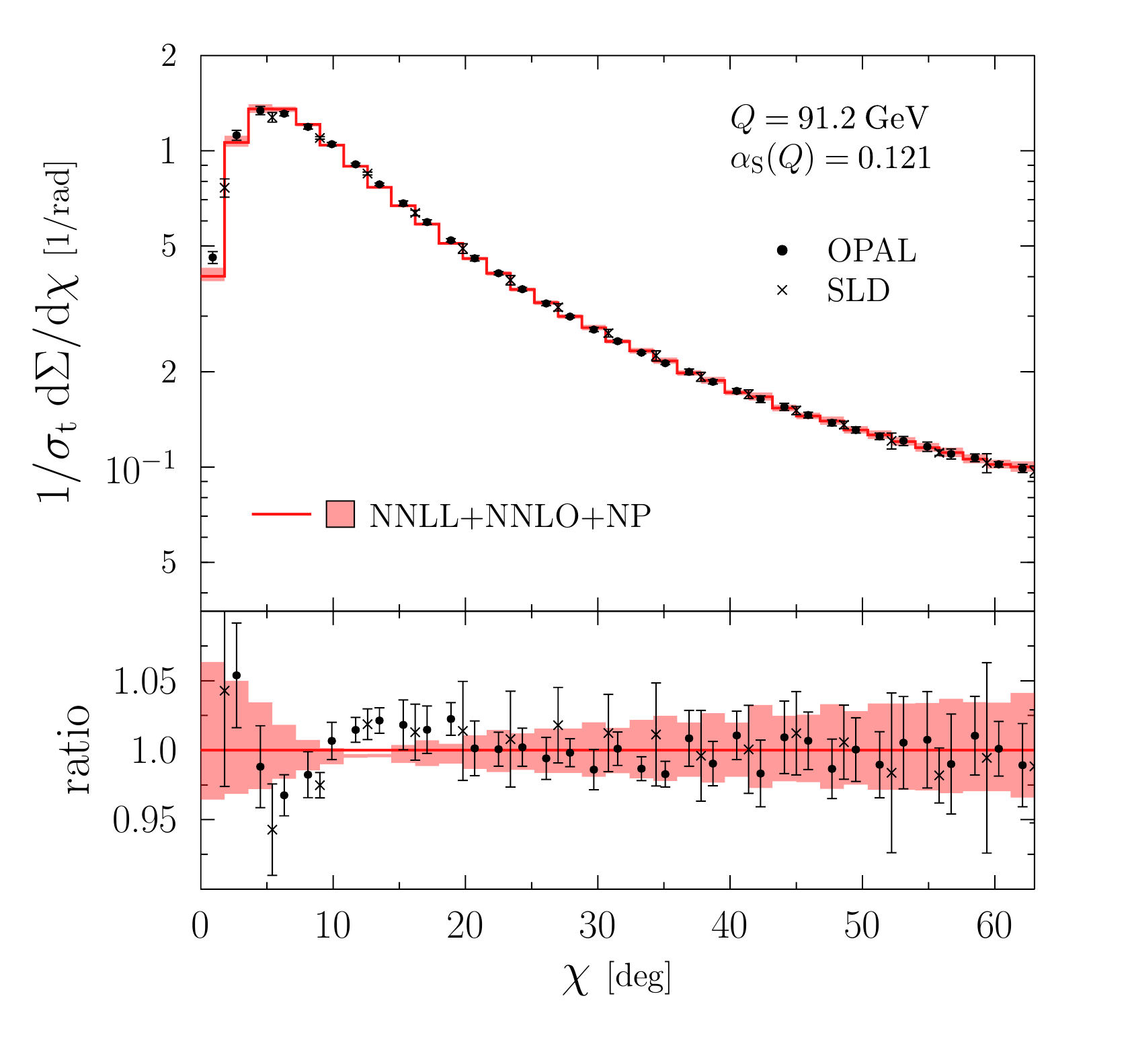}
\caption{NNLL+NNLO matched prediction for EEC. The analytic model of 
\eqn{eq:dEEC-Sudakov-NP} is used to account for hadronization corrections. 
The bottom panel shows the ratio of the data to the matched result. 
The band represents renormalization scale variation in the range $\mu \in [Q/2,2Q]$ 
with three-loop running of $\as$.}
\label{fig:NNLL+NNLO-log-R-fit}
\end{figure}
\Fig{fig:NNLL+NNLO-log-R-fit} shows the comparison of the best fit NNLL+NNLO result 
to the measured data. We again observe that the measurement is very well described by the 
theoretical prediction and, in particular, the impact of the NNLO correction is clearly 
visible in the medium $\chi$ range, where the agreement between the data and the prediction 
is now excellent. The systematic deviation which is present in the NNLL+NLO predictions 
in this range is completely erased when the NNLO correction is taken into account. At the 
same time the best fit value of $\as(M_Z)$ is shifted by about $-6\%$. We conclude that the 
inclusion of the fixed-order NNLO correction is essential for a precise determination of 
$\as$ from EEC.

Finally, the three-parameter fits show that in this approach to hadronization corrections, 
the non-perturbative parameter $a_1$ is more important than $a_2$. As stressed already in 
\refr{deFlorian:2004mp}, this indicates that the parametrization in \eqn{eq:dEEC-Sudakov-NP} 
is not able to fully describe the non-perturbative corrections, especially at medium and 
large $\chi$. Hence, part of the hadronization effects are absorbed into the strong coupling. 
This is also apparent from the very strong anti-correlation in the fits between $\as$ and 
the non-perturbative parameter $a_2$. Thus, it would be very interesting to repeat our analysis 
with hadronization corrections extracted from data by comparison to Monte Carlo simulations. 
The results of such an analysis will appear elsewhere \cite{inprep:2017xx}.


\section{Conclusions}
\label{sec:concl}

In this paper we presented precise QCD predictions for the energy-energy correlation 
in $e^+e^-$ collisions. Our computation includes fixed-order perturbative corrections 
up to NNLO accuracy, as well as a resummation of the logarithmically enhanced terms 
in the back-to-back region at NNLL accuracy. In order to obtain a description which 
incorporates the complete perturbative knowledge about the observable and is valid 
over a wide kinematical range, the fixed-order and resummed predictions must be matched. 
We have implemented this matching in the $R$ scheme at NNLL+NLO and also, for 
the first time, in the log-$R$ scheme at both NNLL+NLO and NNLL+NNLO accuracy. All of 
our matched results satisfy the physical requirement that the EEC distribution should 
vanish as $\chi \to 0$.

We also presented perturbative predictions at NNLL+NLO and NNLL+NNLO accuracy and 
compared these to precise OPAL and SLD data. In particular, we have performed a fit 
of our results to the data with the strong coupling $\as$ as a free parameter. Using 
an analytic model to account for hadronization corrections, we obtain a very good 
description of the data down to the smallest measured angles. We observe that the 
inclusion of the NNLO corrections has a significant impact on the extracted value of 
$\as(M_Z)$, shifting the best fit value by around $-6\%$ compared to the NNLL+NLO 
computation. Hence, the inclusion of these corrections in a precise measurement of 
$\as$ from EEC is mandatory. 

Using our most accurate NNLL+NNLO theoretical prediction and \eqn{eq:dEEC-Sudakov-NP} 
to model the non-perturbative contributions, we obtain our best fit value of 
$\as(M_Z) = 0.121^{+0.001}_{-0.003}$ which is compatible with the world average within 
uncertainties. It would be very interesting to perform a more comprehensive 
phenomenological analysis and a precise measurement of $\as$ from EEC, using modern Monte 
Carlo tools to extract the hadronization corrections from data. This work is in progress.


\section*{Acknowledgments}

We gratefully acknowledge A.~Verbytskyi and S.~Kluth for helpful discussions and 
comments on the manuscript. We are also grateful to M.~Grazzini, P.~Monni, Z.~Sz\H or 
and Z.~Tr\'ocs\'anyi for useful discussions and suggestions. This research was supported 
by grant K 125105 of the National Research, Development and Innovation Office in Hungary. 
ZT was supported by the \'UNKP-17-3 New National Excellence Program of the Ministry of Human 
Capacities of Hungary. AK acknowledges financial support from the Postdoctoral Fellowship 
program of the Hungarian Academy of Sciences.


\appendix


\section{The $\bar{\mathcal A}_{\mathrm{res.}}$, $\bar{\mathcal B}_{\mathrm{res.}}$ 
and $\bar{\mathcal C}_{\mathrm{res.}}$ coefficients} 
\label{appx:RES-EXP}

Recall that $\bar{\mathcal A}_{\mathrm{res.}}$, $\bar{\mathcal B}_{\mathrm{res.}}$ 
and $\bar{\mathcal C}_{\mathrm{res.}}$ are the coefficients obtained by expanding 
the resummed component of \eqn{eq:EECcos-log-R-match}, 
$
\frac{1}{H(\as(\mu))}
	\Big[\frac{1}{\tsigT} \widetilde{\Sigma}(\chi,\mu)\Big]_{\mathrm{res.}}
$, in a power series in $\as$:
\beq
\bsp
\frac{1}{H(\as(\mu))}
	\bigg[\frac{1}{\tsigT} \widetilde{\Sigma}(\chi,\mu)\bigg]_{\mathrm{res.}} &=
	1
	+
	\frac{\as(\mu)}{2\pi} \bar{\mathcal A}_{\mathrm{res.}}(\chi,\mu) 
	+
	\left(\frac{\as(\mu)}{2\pi}\right)^2 \bar{\mathcal B}_{\mathrm{res.}}(\chi,\mu) 
\\&\qquad
	+
	\left(\frac{\as(\mu)}{2\pi}\right)^3 \bar{\mathcal C}_{\mathrm{res.}}(\chi,\mu)
	+
	\Oa{4}\,.
\label{eq:app-EECcos-res-exp}
\esp
\eeq
These coefficients read (recall $y = \sin^2 \frac{\chi}{2}$)
\bal
\bar{\mathcal A}_{\mathrm{res.}}(\chi,\mu) &= \frac{1}{4}\bigg\{
	-A^{(1)} \ln^2(y) 
	+ 2 (B^{(1)}+A^{(1)} y)\ln(y) 
	-2 (A^{(1)}+B^{(1)}) y
\bigg\}\,,
\\
\bar{\mathcal B}_{\mathrm{res.}}(\chi,\mu) &= \frac{1}{16}\bigg\{
	\frac{(A^{(1)})^2}{2}\ln^4(y) 
	+ \bigg[\frac{4 A^{(1)} \beta_0}{3}-2 A^{(1)} B^{(1)}-2 (A^{(1)})^2 y\bigg]\ln^3(y) 
\\ &	
	+ \bigg[-2 A^{(2)}-2 \beta_0 B^{(1)}+2 (B^{(1)})^2+\bigg(6 (A^{(1)})^2-4 A^{(1)} 
	\beta_0+6 A^{(1)} B^{(1)}\bigg) y
\nt \\ &
	-2 A^{(1)} \beta_0 \ln\xiRsq\bigg]\ln^2(y) + \bigg[4 B^{(2)}+8 
	(A^{(1)})^2 \zeta_3+4 \beta_0 B^{(1)} \ln\xiRsq
\nt \\ &	
	+y \bigg(-12 (A^{(1)})^2+4 A^{(2)}+8 A^{(1)} \beta_0-12 A^{(1)} B^{(1)}+4 \beta_0 B^{(1)}
\nt \\ &	
	-4 (B^{(1)})^2+4 A^{(1)} \beta_0 \ln\xiRsq\bigg)\bigg]\ln(y) 
	+ \bigg[\frac{16 A^{(1)} \beta_0 \zeta_3}{3}-8 A^{(1)} B^{(1)} \zeta_3
\nt \\ &	
	+y \bigg(12 (A^{(1)})^2-4 A^{(2)}-8 A^{(1)} \beta_0+12 A^{(1)} B^{(1)}-4 \beta_0 B^{(1)}
	+4 (B^{(1)})^2-4 B^{(2)}
\nt \\ &		
	-8 (A^{(1)})^2 \zeta_3+(-4 A^{(1)} \beta_0
	-4 \beta_0 B^{(1)}) \ln\xiRsq\bigg)\bigg]
\bigg\}\,,
\nt\\
\bar{\mathcal C}_{\mathrm{res.}}(\chi,\mu) &= \frac{1}{64}\bigg\{
	-\frac{(A^{(1)})^3}{6}\ln^6(y) + \bigg[-\frac{4 (A^{(1)})^2 \beta_0}{3}
	+(A^{(1)})^2 B^{(1)}+(A^{(1)})^3 y\bigg]\ln^5(y) 
\\ &	
	+ \bigg[2 A^{(1)} A^{(2)}-2 A^{(1)} \beta_0^2+\frac{14 A^{(1)} \beta_0 B^{(1)}}{3}
	-2 A^{(1)} (B^{(1)})^2+\bigg(-5 (A^{(1)})^3
\nt \\ &	
	+\frac{20 (A^{(1)})^2 \beta_0}{3}-5 (A^{(1)})^2 B^{(1)}\bigg) y
	+2 (A^{(1)})^2 \beta_0 \ln\xiRsq\bigg]\ln^4(y) 
\nt \\ &		
	+ \bigg[\frac{16 A^{(2)} \beta_0}{3}+\frac{8 A^{(1)} \beta_1}{3}-4 A^{(2)} B^{(1)}
	+\frac{8 \beta_0^2 B^{(1)}}{3}-4 \beta_0 (B^{(1)})^2+\frac{4 (B^{(1)})^3}{3}
\nt \\ &	
	-4 A^{(1)} B^{(2)}-\frac{40 (A^{(1)})^3 \zeta_3}{3}+\bigg(\frac{16 A^{(1)} \beta_0^2}{3}
	-8 A^{(1)} \beta_0 B^{(1)}\bigg) \ln\xiRsq
\nt \\ &	
	+y \bigg(20 (A^{(1)})^3-8 A^{(1)} A^{(2)}-\frac{80 (A^{(1)})^2 \beta_0}{3}
	+8 A^{(1)} \beta_0^2+20 (A^{(1)})^2 B^{(1)}
\nt \\ &	
	-\frac{56 A^{(1)} \beta_0 B^{(1)}}{3}+8 A^{(1)} (B^{(1)})^2
	-8 (A^{(1)})^2 \beta_0 \ln\xiRsq\bigg)\bigg]\ln^3(y)
\nt \\ &	
	 + \bigg[-4 A^{(3)}-4 \beta_1 B^{(1)}-8 \beta_0 B^{(2)}+8 B^{(1)} B^{(2)}
	 -\frac{160}{3} (A^{(1)})^2 \beta_0 \zeta_3+40 (A^{(1)})^2 B^{(1)} \zeta_3
\nt \\ &	
	 +\bigg(-8 A^{(2)} \beta_0-4 A^{(1)} \beta_1-8 \beta_0^2 B^{(1)}
	 +8 \beta_0 (B^{(1)})^2\bigg) \ln\xiRsq
	 -4 A^{(1)} \beta_0^2 \ln^2\xiRsq
\nt \\ &	
	 +y \bigg(-60 (A^{(1)})^3+24 A^{(1)} A^{(2)}+80 (A^{(1)})^2 \beta_0
	 -16 A^{(2)} \beta_0-24 A^{(1)} \beta_0^2-8 A^{(1)} \beta_1
\nt \\ &	
	 -60 (A^{(1)})^2 B^{(1)}+12 A^{(2)} B^{(1)}+56 A^{(1)} \beta_0 B^{(1)}
	 -8 \beta_0^2 B^{(1)}-24 A^{(1)} (B^{(1)})^2
\nt \\ &	
	 +12 \beta_0 (B^{(1)})^2-4 (B^{(1)})^3+12 A^{(1)} B^{(2)}+40 (A^{(1)})^3 \zeta_3
	 +\bigg(24 (A^{(1)})^2 \beta_0-16 A^{(1)} \beta_0^2
\nt \\ &	
	 +24 A^{(1)} \beta_0 B^{(1)}\bigg) \ln\xiRsq\bigg)\bigg]\ln^2(y) 
	 + \bigg[32 A^{(1)} A^{(2)} \zeta_3-32 A^{(1)} \beta_0^2 \zeta_3
	 +\frac{224}{3} A^{(1)} \beta_0 B^{(1)} \zeta_3
\nt \\ &		 
	 -32 A^{(1)} (B^{(1)})^2 \zeta_3-48 (A^{(1)})^3 \zeta_5
	 +32 (A^{(1)})^2 \beta_0 \zeta_3 \ln\xiRsq
	 +y \bigg(120 (A^{(1)})^3-48 A^{(1)} A^{(2)}
\nt \\ & 	 
	 +8 A^{(3)}-160 (A^{(1)})^2 \beta_0+32 A^{(2)} \beta_0
	 +48 A^{(1)} \beta_0^2+16 A^{(1)} \beta_1+120 (A^{(1)})^2 B^{(1)}
\nt \\ &	 	 	 
	 -24 A^{(2)} B^{(1)}-112 A^{(1)} \beta_0 B^{(1)}+16 \beta_0^2 B^{(1)}+8 \beta_1 B^{(1)}
 	 +48 A^{(1)} (B^{(1)})^2-24 \beta_0 (B^{(1)})^2
\nt \\ &		 
	 +8 (B^{(1)})^3-24 A^{(1)} B^{(2)}
	 +16 \beta_0 B^{(2)}-16 B^{(1)} B^{(2)}
	 -80 (A^{(1)})^3 \zeta_3+\frac{320}{3} (A^{(1)})^2 \beta_0 \zeta_3
\nt \\ &	 	 
	 -80 (A^{(1)})^2 B^{(1)} \zeta_3+\bigg(-48 (A^{(1)})^2 \beta_0+16 A^{(2)} \beta_0
	 +32 A^{(1)} \beta_0^2+8 A^{(1)} \beta_1-48 A^{(1)} \beta_0 B^{(1)}
\nt \\ &	
	 +16 \beta_0^2 B^{(1)}-16 \beta_0 (B^{(1)})^2\bigg) \ln\xiRsq
	 +8 A^{(1)} \beta_0^2 \ln^2\xiRsq\bigg)\bigg]\ln(y) 
	 + \bigg[\frac{64 A^{(2)} \beta_0 \zeta_3}{3}
\nt \\ &		 
	 +\frac{32 A^{(1)} \beta_1 \zeta_3}{3}-16 A^{(2)} B^{(1)} \zeta_3
	 +\frac{32}{3} \beta_0^2 B^{(1)} \zeta_3-16 \beta_0 (B^{(1)})^2 \zeta_3
	 +\frac{16 (B^{(1)})^3 \zeta_3}{3}
\nt \\ &	 	 
	 -16 A^{(1)} B^{(2)} \zeta_3-\frac{80 (A^{(1)})^3 \zeta_3^2}{3}
	 -64 (A^{(1)})^2 \beta_0 \zeta_5
	 +48 (A^{(1)})^2 B^{(1)} \zeta_5+\bigg(\frac{64}{3} A^{(1)} \beta_0^2 \zeta_3
\nt \\ &	
	 -32 A^{(1)} \beta_0 B^{(1)} \zeta_3\bigg) \ln\xiRsq
	 +y \bigg(-120 (A^{(1)})^3+48 A^{(1)} A^{(2)}-8 A^{(3)}+160 (A^{(1)})^2 \beta_0
\nt \\ &	
	 -32 A^{(2)} \beta_0-48 A^{(1)} \beta_0^2 
	 -16 A^{(1)} \beta_1-120 (A^{(1)})^2 B^{(1)}+24 A^{(2)} B^{(1)}
	 +112 A^{(1)} \beta_0 B^{(1)}
\nt \\ &		 
	 -16 \beta_0^2 B^{(1)}-8 \beta_1 B^{(1)}
	 -48 A^{(1)} (B^{(1)})^2+24 \beta_0 (B^{(1)})^2-8 (B^{(1)})^3+24 A^{(1)} B^{(2)}
\nt \\ &	 
	 -16 \beta_0 B^{(2)}+16 B^{(1)} B^{(2)}+80 (A^{(1)})^3 \zeta_3-32 A^{(1)} A^{(2)} \zeta_3
	 -\frac{320}{3} (A^{(1)})^2 \beta_0 \zeta_3+32 A^{(1)} \beta_0^2 \zeta_3
\nt \\ &	 
	 +80 (A^{(1)})^2 B^{(1)} \zeta_3
	 -\frac{224}{3} A^{(1)} \beta_0 B^{(1)} \zeta_3+32 A^{(1)} (B^{(1)})^2 \zeta_3
	 +48 (A^{(1)})^3 \zeta_5
\nt \\ & 	 	 
	 +\bigg(48 (A^{(1)})^2 \beta_0
	 -16 A^{(2)} \beta_0-32 A^{(1)} \beta_0^2-8 A^{(1)} \beta_1+48 A^{(1)} \beta_0 B^{(1)}
	 -16 \beta_0^2 B^{(1)}
\nt \\&	 	 
	 +16 \beta_0 (B^{(1)})^2
	 -32 (A^{(1)})^2 \beta_0 \zeta_3\bigg) \ln\xiRsq
	 -8 A^{(1)} \beta_0^2 \ln^2\xiRsq\bigg)\bigg]
\bigg\}\,.
\nt
\eal
%




\providecommand{\href}[2]{#2}\begingroup\raggedright\endgroup


\end{document}